\documentclass[11pt]{article}
\pdfoutput=1

\usepackage{graphics, color,soul}
\usepackage{graphicx}
\usepackage{amssymb}

\usepackage{lmodern,mathtools}

\usepackage{booktabs}
\usepackage[english]{babel}
\usepackage{amsmath,amssymb,amsbsy,amstext, amsthm, simplewick}
\usepackage{hyperref}
\usepackage{tikz}

\usetikzlibrary{decorations.pathmorphing}
\tikzset{snake it/.style={decorate, decoration=snake}}

\usepackage{amsfonts}
\usepackage{amssymb}
\usepackage{upgreek}
\usepackage{simplewick}
 \usepackage{exscale,relsize}
\usepackage{mathtools}
\usepackage{comment}

\usepackage[margin=1cm,labelfont={sf,bf,scriptsize},textfont={sf,scriptsize}]{caption}

\usepackage{colortbl}
\definecolor{lightgreen}{cmyk}{0.2, 0, 0.2, 0.2}
\definecolor{lightgray}{cmyk}{0.1,0.2,0,0.1}
\definecolor{lightgray2}{cmyk}{0.1,0.1,0,0.1}

\makeatletter
\newlength{\apb@width}
\newcommand{\autoparbox}[2][c]{\settowidth{\apb@width}{#2}\parbox[#1]{\apb@width}{#2}}

\makeatother

\numberwithin{equation}{section}

\def\beq{\begin{equation}}
\def\eeq{\end{equation}}

\def\bea{\begin{eqnarray}}
\def\eea{\end{eqnarray}}

\def\beq{\begin{equation}}
\def\eeq{\end{equation}}
\def\be{\begin{equation}}
\def\ee{\end{equation}}
\def\bea{\begin{eqnarray}}
\def\eea{\end{eqnarray}}

\def\0{{\vec{0}}}

\DeclareRobustCommand{\SkipTocEntry}[4]{}

\def\beq{\begin{equation}}
\def\eeq{\end{equation}}

\def\ba#1\ea{\begin{align}#1\end{align}}
\def\bg#1\eg{\begin{gather}#1\end{gather}}
\newcommand{\bseq}{\begin{subequations}}
\newcommand{\eseq}{\end{subequations}}

\DeclareSymbolFont{extraup}{U}{zavm}{m}{n}
\DeclareMathSymbol{\varheart}{\mathalpha}{extraup}{86}
\DeclareMathSymbol{\vardiamond}{\mathalpha}{extraup}{87}

\def\({\left(}
\def\){\right)}
\def\[{\left[}
\def\]{\right]}

\begin{document}

\begin{titlepage}
{~~~~~~~~~~~~~~~~~~~~~~~~~~~~~~~~~~~~
~~~~~~~~~~~~~~~~~~~~~~~~~~~~~~~~~~
~~~~~~~~~~~ \footnotesize{SLAC-PUB-16260,SU/ITP-15/04,NSF-KITP-15-047}} 
\setcounter{page}{1} \baselineskip=15.5pt \thispagestyle{empty}

\vbox{\baselineskip14pt
}
{~~~~~~~~~~~~~~~~~~~~~~~~~~~~~~~~~~~~
~~~~~~~~~~~~~~~~~~~~~~~~~~~~~~~~~~
~~~~~~~~~~~ }

\bigskip\

\vspace{2cm}
\begin{center}

{\fontsize{19}{36}\selectfont  \sc
Longitudinal nonlocality in the\\
 string S-matrix}
\end{center}

\vspace{0.6cm}

\begin{center}
{\fontsize{13}{30}\selectfont  Matthew Dodelson$^{1,2}$ and Eva Silverstein$^{1,3,4}$}
\end{center}


\begin{center}
\vskip 8pt

\textsl{
\emph{$^1$Stanford Institute for Theoretical Physics, Stanford University, Stanford, CA 94306}}

\vskip 7pt
\textsl{ \emph{$^2$Kavli Institute for Theoretical Physics, University of California, Santa Barbara, CA 93106}}

\vskip 7pt
\textsl{ \emph{$^3$SLAC National Accelerator Laboratory, 2575 Sand Hill, Menlo Park, CA 94025}}

\vskip 7pt
\textsl{ \emph{$^4$Kavli Institute for Particle Astrophysics and Cosmology, Stanford, CA 94025}}

\end{center}

\vspace{0.5cm}
\hrule \vspace{0.1cm}
{ \noindent \textbf{Abstract} \\[0.01cm]
We analyze four and five-point tree-level open string S-matrix amplitudes in the Regge limit, exhibiting some basic features which indicate longitudinal nonlocality, as suggested by light cone gauge calculations of string spreading. Using wavepackets to localize the asymptotic states, we compute the peak trajectories followed by the incoming and outgoing strings, determined by the phases in the amplitudes.  These trajectories trace back in all dimensions such that the incoming strings deflect directly into corresponding outgoing ones, as expected from a Reggeon analysis.  Bremsstrahlung radiation at five points emerges from the deflection point, corroborating this picture.    An explicit solution for the intermediate state produced at four points in the $s$-channel exists, with endpoints precisely following the corresponding geometry and a periodicity which matches the series of time delays predicted by the amplitude.  We find a nonzero peak impact parameter for this process, and show that it admits an interpretation in terms of longitudinal-spreading induced string joining, at the scale expected from light cone calculations, and does not appear to admit a straightforward interpretation purely in terms of the well-established transverse spreading.  At five points, we exhibit a regime with advanced emission of one of the deflected outgoing strings.  This strongly suggests early interaction induced by longitudinal nonlocality.  In a companion paper, we apply string spreading to horizon dynamics.   

\vspace{0.3cm}
 \hrule

\vspace{0.6cm}}
\end{titlepage}

\tableofcontents
\section{Introduction}

It has long been understood that string theory provides a strong candidate for an ultraviolet completion of gravity.  In perturbative limits of the theory, the S-matrix amplitudes are finite aside from infrared divergences of physical origin.  Related phenomena such as singularity resolution and smooth topology changing transitions arise already in the perturbative -- even classical -- theory.  Duality conjectures and the mathematics supporting them provide strong evidence that the perturbative amplitudes fit concretely into a complete, non-perturbative theory of quantum gravity.   

However, essential questions remain even in the perturbative theory.  In this paper, we will be concerned with a longstanding question about the degree of {\it longitudinal spreading} that plays a role in string interactions.      
Exploiting ideas from hadron physics, Susskind \cite{lennyspreading}\ computed the variance of the transverse and longitudinal embedding coordinates of the string, using the explicit light cone single-string ground-state wavefunction.   We take the longitudinal direction $X^+$ along the direction of relative motion of the strings -- more precisely, it is defined in the brick wall frame \cite{bpst}
\be\label{brickwall}
p_{\perp,r}=\pm \frac{k_\perp}{2}
\ee
where the momentum transfer $k_\perp$ is divided equally between the incoming and outgoing strings, which are indexed by $r$, with transverse momentum $p_{\perp, r}$.
Given this, one obtains 
\begin{align}
\langle (\Delta X_\perp)^2\rangle &= \alpha' \sum_{n=1}^{n_{\text{max}}}\frac{1}{n} =\alpha'\log \frac{n_{\text{max}}}{n_0}+ O(1/n_{\text{max}}) \notag \\
\langle (\Delta X^+)^2\rangle &\sim \frac{1}{(p^-)^2} \sum_{n=1}^{n_{\text{max}}}n \sim \frac{{n_{\text{max}}^2}}{(p^-)^2},\label{spreadapprox}
\end{align}
with $n_0$ a constant and $n_{\text{max}}$ determined by the light cone time resolution of the detector; for string scattering with $s\gg -t\gg 1/\alpha'$ this is given by 
\be\label{nmaxintro}
n_{\text{max}} \sim \frac{ s}{k_\perp^2}\sim -\frac{s}{t}.
\ee
We have worked in the string ground state; these expressions are valid for sufficiently small mass.
In \cite{usBH}, we review this effect in more detail at the level of the light cone calculations in \cite{lennyspreading}\cite{bpst}, clarifying its consistency with the underlying Lorentz symmetry of the theory.  

The transverse spreading $\langle (\Delta X_\perp)^2\rangle$ has been relatively well established, via the impact parameter transform of forward scattering at four points, and in calculations such as \cite{bpst}\ which explicitly manifest the cutoff $n_{\text{max}}$ as we will discuss further below.  Intuitively, it is more straightforward to measure the transverse distribution of string via head-on scattering than it is to tease out the longitudinal extent of the string.  At the level of the light cone gauge calculations, the two go together:  a constraint directly relates the first and second lines of (\ref{spreadapprox}).  

The longitudinal spreading, if not a gauge artifact,  has important consequences beyond  flat space string amplitudes.  It was originally applied in \cite{lennyspreading}\ to black hole physics, realizing the idea of a stretched horizon in a concrete way.  In that work, it was assumed that effective field theory does not break down for an infalling observer.\footnote{In general, the validity and implications of this effect have remained rather mysterious despite much interesting work \cite{locality}\cite{JoeBHcomp}\cite{ggm}.}   However, in a companion paper \cite{usBH}\ we show that the longitudinal spreading as computed in light cone gauge implies    a breakdown of effective field theory and some `drama' for a class of probes falling into the black hole long after an early infalling string \cite{backdraft}.  This provides a concrete approach to the longstanding problem recently sharpened in \cite{firewalls}.      

Given these motivations, it is important to determine whether the putative longitudinal spreading plays a clear role in gauge-invariant observables such as S-matrix amplitudes.  In this paper we will show that assuming the limited extent of the transverse spreading described above -- which is already supported by substantial evidence -- certain features of string S-matrix amplitudes require longitudinal nonlocality.        

Our basic strategy employs the phases in tree-level open string amplitudes at four and five points and the peak trajectories they imply for the incoming and outgoing strings (localized using wavepackets).  We find in some cases that longitudinal spreading is required to interpret the results consistently with causality.  It is particularly useful to keep track of the apparent time delay or advance of each outgoing string, relative to a putative center of mass collision.\footnote{We thank S. Giddings for suggesting time advances as a probe of longitudinal spreading, as well as pointing us to \cite{juancausality}.}  A time advance -- the emergence of an outgoing string before the would-be center of mass collision -- would immediately require longitudinal nonlocality.  In ordinary quantum mechanics, scattering off of a repulsive potential produces a time advance, while attractive potentials (even if extended in the direction of relative motion) lead to time delays.    

At four points \cite{Veneziano}, two of the three open string diagrams exhibit net time delays, and one has neither a delay nor an advance. The latter, marginal case motivates a careful study at five points to see which way the net time shift goes with the inclusion of an additional probe. The four point diagrams with net delays may simply indicate that string interactions are attractive.  Even in that case, we find a peak impact parameter at nonzero scattering angle which is not explained purely by the transverse spreading $\langle (\Delta X_\perp)^2\rangle \sim \alpha'  \log n_{\text{max}} $ just reviewed.   A simple intermediate string solution along the lines of \cite{yoyo1}-\cite{yoyo3} captures the impact parameter and time delay in a simple and explicit way, and does not admit a purely transverse description given the distribution reviewed above.  

Moving to five points, we find a net time advance in a generalization of the diagram which at four points had no time shift, perturbing it with an additional outgoing leg of energy much smaller than that of the incoming strings A and B.   
Working in a Regge limit where an outgoing string (labeled String 1) emerges from a particular incoming string (String A), we find that the peak trajectories imply an apparent time advance for String 1.  Again conditioned on the standard transverse string spreading -- which implies a penalty for emission of 1 at a transverse distance from String A -- we show that the apparent advance is real, so that an early interaction is required by causality.  This provides strong evidence for longitudinal nonlocality in the string S-matrix.
      
An additional motivation for the five-point function analysis was to search for a signal of early interaction in Bremsstrahlung radiation.  Although this does not arise in the regimes we have analyzed thus far, we find that the peak trajectory for the outgoing radiation is precisely consistent with emission at sharp turning points of the $s$-channel string solutions which provide a simple and quantitative fit to the amplitudes.

\section{Four point scattering: time shifts and peak impact parameters}
\label{fourpoints}

S-matrix amplitudes are well-defined observables in string theory, but finite-time Green's functions are not.  As such, the detailed evolution between asymptotic regions is nontrivial to extract, requiring additional probes beyond the simplest $2\to 2$ scattering process.   By eventually going to five points, we will find a relatively simple derivation of longitudinal nonlocality -- given the standard scale of transverse spreading.   In this section, we will warm up with four-point functions (the Veneziano and Virasoro-Shapiro amplitudes), analyzing their phases, time shifts (delays or advances relative to a localized center of mass collision), and peak impact parameters.  Along the way, we will present simple models of intermediate string configurations which match these behaviors quantitatively, while noting some remaining open questions.  The calculations in this section will lay the groundwork for an analysis of open string five-tachyon amplitudes starting in Section \ref{fiverad}.

Before getting to the concrete calculations, let us describe the results briefly.
The phases of the four-point string amplitudes will imply time shifts that are  net time delays in some cases, as well as a marginal case with zero delay or advance.  This latter case will prove interesting when generalized to the five-point level. 

In ordinary quantum mechanics, time delays occur for attractive potentials, while advances occur for repulsive ones.    The net time delays in certain four-point diagrams may simply reflect the attractive nature of the string interactions in these processes.
If strings have some nonzero longitudinal size, then one may naively expect to find a time advance in the $2\to2$ scattering amplitude for backscattering. Note, however, that this process only occurs if the interaction is repulsive, as would be the case for the scattering of rigid rods. Strings, on the other hand, are attractive by nature; the tension of a string pulls the ends towards the center.
 It is therefore possible that the strings begin to interact before the center of masses collide, without giving a time advance in backscattering kinematics.

Even at the four point function level, a simple analysis convolving the standard amplitudes with wavepackets will expose a curious feature, a nonzero peak value of the impact parameter for certain worldsheet topologies.   After deriving that feature, we will return to a possible interpretation in terms of longitudinal spreading.

\subsection{Preparing the wavepackets}\label{wavepackets}
\indent Consider the four-point scattering amplitude $A(s,t)$, where the Mandelstam invariants are defined as usual by 
\begin{align}
s=-(k_A+k_B)^2\hspace{10 mm} t=-(k_A+k_1)^2\hspace{10 mm}u=-(k_A+k_3)^2.
\end{align}
The initial strings are labeled by A and B, and the final strings are labeled by 1 and 3 (anticipating the addition of an outgoing radiation mode 2 at five points in the following sections). Fixing the center of mass frame and taking all the strings to be highly relativistic, the momenta for the scattering process are 
\begin{align}
k_A&=(E,E,0)\notag\\
k_B&=(E,-E,0)\notag\\
k_1&=(-E,-E\cos\theta,-E\sin\theta)\notag\\
k_3&=(-E,E\cos\theta,E\sin\theta)\label{peak1}.
\end{align}
We have chosen the scattering process to occur in the $x-y$ plane, suppressing all other transverse directions for simplicity.\\
\indent Take the initial state to be localized in the transverse and longitudinal coordinates of the initial strings, with strings A and B located at transverse positions $y=b/2$ and $y=-b/2$ respectively,
\begin{align}\label{initial}
|i\rangle&=\int d\tilde{k}_{Ax}\, d\tilde{k}_{Bx}\, d\tilde{k}_+\, d\tilde{k}_- \,e^{-i\tilde{k}_-b/2 }\notag\\
&\hspace{30 mm}\times\exp\left(-\frac{(\tilde{k}_{Ax}-k_{Ax})^2+(\tilde{k}_{Bx}-k_{Bx})^2}{2\sigma_{\text{L}}^2}-\frac{\tilde{k}_{+}^2+\tilde{k}_{-}^2}{2\sigma_\text{T}^2}\right)|\tilde{k}_A,\tilde{k}_B\rangle,
\end{align}
where we have defined 
\begin{align}
\tilde k_{\pm}=\tilde k_{Ay}\pm \tilde k_{By}.
\end{align}
Here the tilded frequencies are $\tilde k^0=\pm|\vec{k}|$ so that we integrate only over states  satisfying the on-shell conditions, with $\tilde k^0$ positive for incoming and negative for outgoing strings.  
Specifically, we have
\bea\label{justsayit}
\tilde k_A &=& k_A + (\delta\tilde k_{Ax}, \delta\tilde k_{Ax}, \tilde k_{Ay})+O(\tilde\delta^2)\\
\tilde k_B &=& k_B + (-\delta\tilde k_{Bx}, \delta\tilde k_{Bx}, \tilde k_{By})+O(\tilde\delta^2)\\
\tilde k_1 &=& k_1 + (\delta\tilde k_{1x_1} , \delta\tilde k_{1x_1}\cos\theta,\delta\tilde k_{1x_1}\sin\theta)+O(\tilde\delta^2).
\eea
where we neglect quadratic deviations in the momenta, as these will be negligible in our Regge (large $E$) limit with small $\sigma$ as we will discuss further below.  To be explicit, we have chosen to use Gaussian wavepackets, although our main conclusions do not depend on this choice.\\
\indent Now let us specify the final state. We are particularly interested in determining when one of the strings, say String 1, emerges. For this purpose it is sufficient to localize String 1 in its longitudinal direction 
\be\label{xonedef}
x_1=x\cos\theta+y\sin\theta.
\ee
In addition, we will allow for a time delay $T$ between when String A is sent in and when String 1 comes out. The corresponding state is 
\begin{align}
\langle f|=\int d\tilde{k}_{1x_1}\, e^{i\tilde{k}_{1x_1}T}\exp\left(-\frac{(\tilde{k}_{1x_1}-k_{1x_1})^2}{2\sigma_\text{L}^2}\right)\langle \tilde{k}_1,k_3|
\end{align}
The scattering amplitude is the overlap of the initial and final states, $\langle f|i\rangle$.\\
\indent The next step is to use the momentum-conserving delta functions to perform three of the integrals. Defining the variables $\delta \tilde{k}=\tilde{k}-k$, energy-momentum conservation is solved when 
\begin{align}
0&=\delta \tilde{k}_{Ax}-\delta \tilde{k}_{Bx}+\delta \tilde{k}_{1x_1}+O(\tilde{\delta}^2)\\
0&=\tilde{k}_{+}+\delta \tilde{k}_{1x_1}\sin\theta\\
0&=\delta \tilde{k}_{Ax}+\delta \tilde{k}_{Bx}+\delta \tilde{k}_{1x_1}\cos\theta .
\end{align}
In the first equation (energy conservation), we have expanded to first order in the variations away from the peak of the wavepacket. As explained in Appendix \ref{gaussian}, this expansion is valid as long as $\sigma^2_{\text{T}}\ll (\alpha' \log s)^{-1}$, which we will assume from now on. Solving in terms of $\delta \tilde{k}_{1x_1}$ and $\tilde{k}_{-}$,
\begin{align}
\tilde{k}_{+}&=-\delta \tilde{k}_{1x_1}\sin\theta+O(\tilde{\delta}^2)\\
\delta \tilde{k}_{Ax}&=-\frac{1}{2}(1+\cos\theta)\delta \tilde{k}_{1x_1}+O(\tilde{\delta}^2)\\
\delta \tilde{k}_{Bx}&=\frac{1}{2}(1-\cos\theta)\delta \tilde{k}_{1x_1}+O(\tilde{\delta}^2).
\end{align}
The amplitude then collapses to
\begin{align}
\langle f|i\rangle=\int d\delta \tilde{k}_{1x_1}\,d \tilde{k}_{-}\, e^{-i\tilde{k}_{-}b/2+i\delta \tilde{k}_{1x_1}T}\exp\left(-\frac{\delta \tilde{k}_{1x_1}^2}{2\sigma_{\text{L},\text{eff}}^2}-\frac{\tilde{k}_{-}^2}{2\sigma_\text{T}^2}\right)A(\tilde{s},\tilde{t}),
\end{align}
where the tilded Mandelstam variables are functions of the integration variables,
\begin{align}
\tilde{s}&=4E^2-4\delta \tilde{k}_{1x_1}E+O(\tilde{\delta}^2)\\
\tilde{t}&=-2E^2(1-\cos\theta)+2E\delta \tilde{k}_{1x_1}(1-\cos\theta)+E\tilde{k}_-\sin\theta+O(\tilde{\delta}^2).\end{align}
The effective longitudinal width $\sigma_{\text{L},\text{eff}}$ is a combination of the longitudinal and transverse widths whose precise form will not be important for us.\\
\indent Suppose now that $A$ contains a factor that oscillates rapidly with energy, such that it is possible to write 
\begin{align}
A(s,t)=\exp(i\delta(s,t))A_{\text{slow}}(s,t),
\end{align}
where $A_{\text{slow}}(s,t)$ does not oscillate over the support of $\psi_{E}$. The integral is largest when the phase is stationary at the peak of the wavepackets, which occurs when 
\begin{align}
b&=2\frac{\partial\delta}{\partial \tilde{k}_-}=2E\sin\theta\frac{\partial\delta}{\partial t}\\
T&=-\frac{\partial \delta}{\partial \delta \tilde{k}_{1x_1}}=4E\frac{\partial\delta}{\partial s}-2E(1-\cos\theta)\frac{\partial\delta}{\partial t}.
\end{align}
This yields a time delay for String 1 if the derivative of the phase with respect to $\delta \tilde{k}_{1x}$ is negative, and a time advance otherwise. The peak impact parameter is similarly determined in terms of the derivative of the phase with respect to $\tilde{k}_-$.
\subsection{Closed strings}
Let us start by analyzing the Virasoro-Shapiro amplitude in the Regge limit $s\gg t$. We will work with tachyonic strings at energies much higher than the string scale, so that they are effectively massless. The amplitude is
\begin{align}
A(s,t)&=\frac{g_{\text{c}}^2}{\alpha'}\frac{\Gamma(-1-\alpha' s/4)\Gamma(-1-\alpha' t/4)\Gamma(-1-\alpha' u/4)}{\Gamma(2+\alpha' s/4)\Gamma(2+\alpha' t/4)\Gamma(2+\alpha' u/4)},
\end{align}
where $s+t+u=-16/\alpha' $. We have suppressed the standard $i\epsilon$ prescription, with $s\to s+i\epsilon$, etc. This has poles at integer-spaced energies $\alpha' s=n-1$ for $n\ge 0$, corresponding to the production of massive on-shell strings in the $s$-channel. To isolate the effects of these poles, we can use the identity
\begin{align}
\Gamma(z)\Gamma(1-z)=\frac{\pi}{\sin(\pi z)}
\end{align}
to convert all the gamma functions with negative argument to gamma functions with positive argument. This gives
\begin{align}
A(s,t)&=\frac{g_{\text{c}}^2}{\alpha'}\frac{\sin(\pi\alpha' u/4)\sin(\pi\alpha' t/4)}{\pi \sin (\pi\alpha' s/4)}\left(\frac{\Gamma(-1-\alpha' t/4)\Gamma(-1-\alpha' u/4)}{\Gamma(2+\alpha' s/4)}\right)^2.
\end{align}
We can now safely take the Regge limit $s\gg t$ in the factor involving the gamma functions. Using Stirling's approximation one finds
\begin{align}
A(s,t)\sim \frac{g_{\text{c}}^2}{\alpha'}\frac{\sin(\pi \alpha' (s+t)/4)\sin(\pi \alpha' t/4)}{ \sin (\pi \alpha' s/4)}\Gamma(-1-\alpha' t/4)^2( \alpha's/4)^{2+\alpha' t/2}\label{virasoro}.
\end{align}
\indent Now we need to address the oscillating prefactor, which has poles whenever a massive string is produced. The key to understanding this factor is that the produced strings are unstable, and will decay in the full quantum theory \cite{GSW}. To take this effect into account, we will include a corresponding decay width $\Gamma$ by shifting $s\to s+2iE\Gamma$. Although we will not need to know the explicit value of $\Gamma$, let us try to give a reasonable estimate. Assume that the intermediate $s$-channel states are semiclassical long strings, with lengths $L\sim \alpha' E$; later in this section we will provide evidence for this claim. The decay of the intermediate string is then an extensive quantity, since the string can split anywhere along its length. It follows that the decay rate is approximately
\begin{align}
\Gamma\sim \frac{g_{\text{c}}^2L}{\alpha'}\sim g_{\text{c}}^2 E.
\end{align}
As long as $g_{\text{c}}>0$, we can now Taylor expand in $e^{-\pi\alpha'\Gamma E}$, 
\begin{align}
\frac{\sin(\pi\alpha' (s+t+2i\Gamma E)/4)\sin(\pi \alpha' t/4)}{ \sin (\pi\alpha' (s+2i\Gamma E)/4)}&\propto(1-e^{-i\pi \alpha' t/2})(1-e^{i\pi \alpha' (s+t)/2}e^{-\pi \Gamma \alpha' E})\notag \\
&\hspace{3 mm}\times \sum_{n=0}^{\infty}e^{i\pi \alpha' ns/2}e^{-\pi\alpha' n\Gamma E}\label{powerseries}.
\end{align}
For weak string coupling, our regime of interest, the higher terms in this expansion are not negligible compared to the first term.  As we will discuss momentarily, this reflects the fact that the intermediate state that is produced by the joining of the incoming strings will oscillate many times before splitting into the two outgoing strings, because of the weakness of the coupling.  The terms in this expansion describe the contributions to the amplitude from different numbers of oscillations prior to the string splitting, similarly to the discussion in \cite{sst}.    
Note that this expression still applies in the formal limit $\Gamma \to 0$. In this limit we need to use the standard prescription $s\to s+i\epsilon$, and formally Taylor expanding in $e^{-\pi\epsilon}$ yields (\ref{powerseries}) as $\Gamma\to 0$.\\
\indent We are now ready to determine the phase shifts in the amplitude. To do this we need to isolate the rapidly oscillating part of the amplitude. When $-t$ is sufficiently large so that it is past the massless pole at $t=0$, the $\Gamma(-1-\alpha' t/4)$ function in (\ref{virasoro}) is a smooth, nonoscillatory function. The same is true for the logarithmic form factor $\exp(\alpha' t\log s/2)$. So the full phase shift comes from the power series in (\ref{powerseries}). A typical term in this series takes one of three forms,
\begin{align}
\textbf{I}: e^{i\pi \alpha' ns/2}\hspace{15 mm}\textbf{II}: e^{i\pi \alpha' (ns-t)/2}\hspace{15 mm}
\textbf{III}:e^{i\pi \alpha'((n+1)s+t)/2},
\end{align}
for $n\ge 0$. The corresponding phase shifts are
\begin{align}
\delta_{\textbf{I}}(n)=\frac{\pi \alpha' ns}{2}\hspace{15 mm}\delta_{\textbf{II}}(n) =\frac{\pi \alpha'(ns-t)}{2}
\hspace{10 mm}\delta_{\textbf{III}}(n)=\frac{\pi \alpha' ((n+1)s+t)}{2}.
\end{align}
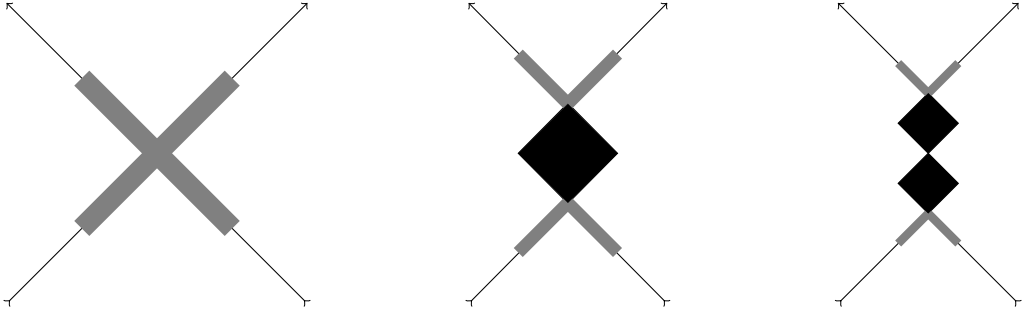
\begin{figure}
\begin{center}
\begin{tikzpicture}[scale=2]
 \draw[>->] (-1,-1) -- (1,1);
 \draw[>->](1,-1) -- (-1,1);
 \fill[color=gray] (-.45,-.55) -- (-.55,-.45) -- (.45,.55) -- (.55,.45) -- (-.45,-.55);
 \fill[color=gray]  (.45,-.55) -- (.55,-.45) -- (-.45,.55) -- (-.55,.45) -- (.45,-.55);
 \end{tikzpicture}\hspace{20 mm}
 \begin{tikzpicture}[scale=2]
 \draw[>->] (-.66,-1) -- (.33,0) -- (-.66,1);
 \draw[>->] (.66,-1) -- (-.33,0) -- (.66,1);
 \fill[color=gray] (-.03,-.30) -- (.03,-.36) -- (-.30,-.69) -- (-.36,-.63) -- (-.03,-.30);
  \fill[color=gray] (.03,-.30) -- (-.03,-.36) -- (.30,-.69) -- (.36,-.63) -- (.03,-.30);
 \fill[color=gray] (-.03,.30) -- (.03,.36) -- (-.30,.69) -- (-.36,.63) -- (-.03,.30);
  \fill[color=gray] (.03,.30) -- (-.03,.36) -- (.30,.69) -- (.36,.63) -- (.03,.30);
  \fill(0,-.33) -- (.33,0) -- (0,.33) -- (-.33,0);
 \end{tikzpicture}
 \hspace{20 mm}
  \begin{tikzpicture}[scale=2]
 \draw[>->] (-.6,-1) -- (.2,-.2) --  (-.2,.2) -- (.6, 1);
 \draw[>->] (.6,-1) -- (-.2,-.2) --  (.2,.2) -- (-.6, 1);
  \fill[color=gray] (.02,-.42) -- (-.02,-.38) -- (-.22,-.58) -- (-.18,-.62) -- (.02,-.42);
  \fill[color=gray] (-.02,-.42) -- (.02,-.38) -- (.22,-.58) -- (.18,-.62) -- (-.02,-.42);
  \fill[color=gray] (.02,.42) -- (-.02,.38) -- (-.22,.58) -- (-.18,.62) -- (.02,.42);
  \fill[color=gray] (-.02,.42) -- (.02,.38) -- (.22,.58) -- (.18,.62) -- (-.02,.42);
    \fill(0,0) -- (.2,-.2) -- (0,-.4) -- (-.2,-.2);
        \fill(0,0) -- (.2,.2) -- (0,.4) -- (-.2,.2);
 \end{tikzpicture}
 \end{center}
 \caption{ \label{oscillating}A schematic diagram of the scattering process for the first few oscillations at $\theta=0$. The region before the center of masses collide that is within the longitudinal spreading radius is shaded in gray. The oscillating long string that is produced by the collision is shaded in black. }
 \end{figure}
\indent Let us now analyze the corresponding time delays in the center of mass frame, starting with the case of forward or backward scattering, $t=0$ \cite{sst}.  The time delays corresponding to the $n$'th phase shifts $\delta_{\textbf{I},\textbf{II}}$ are then
\begin{align}\label{noangles}
4E\frac{\partial \delta_{\textbf{I},\textbf{II}}}{\partial s}(n,\theta=0)=2\pi n\alpha' E.
\end{align}
For $\delta^{\textbf{III}}_n$ the answer is the same, but with $n$ is shifted by one. These time delays are integer multiples of $2\pi \alpha' E$, suggesting a picture of the intermediate string as a classical oscillating string of length $2\pi \alpha' E$. The final strings can be released once in each oscillation, as shown in Figure \ref{oscillating}. Note that the $n$'th phase shift is accompanied by a prefactor
\begin{align}
\exp(-\pi n\alpha' E\Gamma)=\exp\left(-\frac{E\Gamma}{2} \frac{\partial \delta(n)}{\partial s}\right).
\end{align}
Since we have identified $E\partial \delta/\partial s$ as the time delay, this prefactor is reminiscent of a Poisson decay process, where the probability for decaying in time $T$ is $\exp(-\Gamma T)$.\\
\indent Next we turn to the more general case of scattering at nonzero angles in the center of mass frame. Using $t=-2E^2(1-\cos\theta)$, we find 
\begin{align}
T_{\textbf{I}}(n)&=2\pi n\alpha' E\\
T_{\textbf{II}}(n)&=2\pi n\alpha' E+\pi \alpha' E(1-\cos\theta)\label{td1}\\
T_{\textbf{III}}(n)&=2\pi (n+1)\alpha' E-\pi \alpha' E(1-\cos\theta)\label{td2}.
\end{align}
Thus there are angle-dependent corrections to the time delay. \\
\indent Finally, let us compute the impact parameter at which the amplitude is peaked. Consider the phase shift which gives the shortest time delay, $\delta_{\textbf{II},0}$. When the decay rate $\Gamma$ is large, the amplitude is dominated by this phase shift. The peak impact parameter is then
\begin{align}
b=-\pi \alpha' E\sin\theta.
\end{align}
We will discuss the interpretation of this result later in this section.

\subsection{Open strings}
The primary difference between open string and closed string amplitudes is that open strings have different orderings with poles in different channels. At least at four points each ordering can be analyzed independently by introducing Chan-Paton factors. The three independent amplitudes for open string tachyons are 
\begin{align}
A_{st}(s,t)&=\frac{g_{\text{o}}^2}{\alpha'}\frac{\Gamma(-1-\alpha' s)\Gamma(-1-\alpha' t)}{\Gamma(-2-\alpha'(s+t))}\\
A_{su}(s,u)&=\frac{g_{\text{o}}^2}{\alpha'}\frac{\Gamma(-1-\alpha' s)\Gamma(-1-\alpha' u)}{\Gamma(-2-\alpha'(s+u))}\\
A_{tu}(t,u)&=\frac{g_{\text{o}}^2}{\alpha'}\frac{\Gamma(-1-\alpha' t)\Gamma(-1-\alpha' u)}{\Gamma(-2-\alpha'(t+u))}.
\end{align}
The subscripts denote the kinematic invariants in which the amplitudes have poles. Proceeding analogously to the previous section, we find the Regge limits 
\begin{align}\label{fullforms}
A_{st}(s,t)&\sim -\frac{g_{\text{o}}^2}{\alpha'}\frac{\sin(\pi\alpha'(s+t))}{\sin(\pi\alpha' s)}\Gamma(-1-\alpha' t)(\alpha' s)^{1+\alpha' t}\\
A_{su}(s,t)&\sim \frac{g_{\text{o}}^2}{\alpha'}\frac{\sin(\pi \alpha' t)}{\sin(\pi\alpha' s)}\Gamma(-1-\alpha' t)(\alpha' s)^{1+\alpha' t}\\
A_{tu}(s,t)&\sim \frac{g_{\text{o}}^2}{\alpha'}\Gamma(-1-\alpha' t) (\alpha' s)^{1+\alpha' t}.
\end{align}
Shifting $s\to s+2i\Gamma E$ and Taylor expanding as above, we find the time delays
\begin{align}
T_{st}(n)&=8\pi \alpha' (n+1/2\pm 1/2)E\mp 2\pi \alpha' E(1-\cos\theta)\label{timedelayst}\\
T_{su}(n)&=4\pi \alpha'(2n+1) E\pm 2\pi \alpha' E(1-\cos\theta)\label{timedelaysu}.
\end{align}
The third diagram contains no phase shift. The smallest phase shift comes from $A_{st}$, and is of the form $e^{-i\pi \alpha' t}$. As in the previous section, this gives a peak impact parameter
\begin{align}\label{peak4ptsopen}
b=-2\pi \alpha' E\sin\theta .
\end{align}
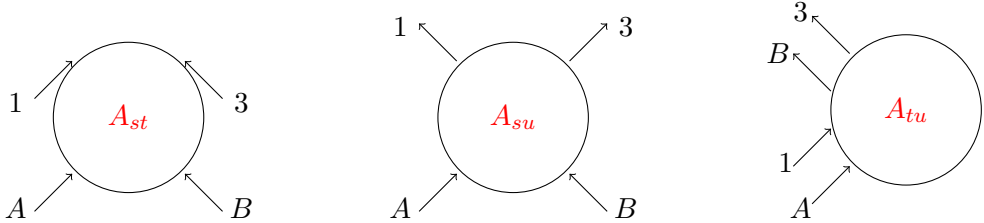
\begin{figure}
\begin{center}
 \begin{tikzpicture}
 \nolinebreak
\draw (0,0) circle (1 cm);
\draw (1.5,-1.2) node {$B$};
\draw (-1.5,-1.2) node {$A$};
\draw (-1.5,.2) node {$1$};
\draw (1.5,.2) node {$3$};
 \draw[color=red] (0,0) node{$A_{st}$}; 
\draw[->]  (-1.25,-1.25) -- (-.75,-.75); 
\draw[->]  (1.25,-1.25) -- (.75,-.75); 
\draw[->]  (-1.25,.25) -- (-.75,.75); 
\draw[->]  (1.25,.25) -- (.75,.75); 
\end{tikzpicture}
 \hspace{13 mm} \begin{tikzpicture}
 \nolinebreak
\draw (0,0) circle (1 cm);
\draw (1.5,-1.2) node {$B$};
\draw (-1.5,-1.2) node {$A$};
\draw (-1.5,1.2) node {$1$};
\draw (1.5,1.2) node {$3$};
\draw[color=red] (0,0) node{$A_{su}$}; 
\draw[->]  (-1.25,-1.25) -- (-.75,-.75); 
\draw[->]  (1.25,-1.25) -- (.75,-.75); 
\draw[<-]  (-1.25,1.25) -- (-.75,.75); 
\draw[<-]  (1.25,1.25) -- (.75,.75); 
\end{tikzpicture}
 \hspace{12 mm} \begin{tikzpicture}
 \nolinebreak
\draw (0,0) circle (1 cm);
\draw (-1.6,-.7) node {$1$};
\draw (-1.4,-1.3) node {$A$};
\draw (-1.7,.75) node {$B$};
\draw (-1.4,1.3) node {$3$};
\draw[color=red]  (0,0) node{$A_{tu}$}; 
\draw[->]  (-1.25,-1.25) -- (-.75,-.75); 
\draw[->]  (-1.5,-.75) -- (-1,-.25); 
\draw[->]  (-1,.25) -- (-1.5,.75); 
\draw[->]  (-.75,.75) -- (-1.25,1.25);
\end{tikzpicture} \end{center}
  \caption{\label{opendiagrams} The three independent open string orderings. We have drawn arrows to signify the spacetime direction of the momenta of each of the strings at $\theta=0$.}
 \end{figure}
\indent We can build some intuition for the phase shifts of each diagram at $\theta=0$ by examining the corresponding orderings on the worldsheet, as displayed in Figure \ref{opendiagrams}. From Figure \ref{opendiagrams}, we see that the ordering for $A_{st}$ corresponds to forward scattering, so the smallest time delay in this ordering should be zero at $\theta=0$. This agrees with (\ref{timedelayst}). \\
\indent The second amplitude $A_{su}$ corresponds to backscattering, as shown in Figure \ref{opendiagrams}. Assuming that a long string is made during the scattering process, it should be possible to backscatter after one full oscillation of the intermediate string, so the shortest time delay should be on the order of $\alpha' E$. This is again in agreement with the result (\ref{timedelaysu}).

 Finally, the third amplitude $A_{tu}$ does not have poles in the $s$-channel, so it is impossible to make an on-shell intermediate resonance, which explains why the phase is trivial. The manifestation of this fact in the worldsheet diagram in Figure \ref{opendiagrams} is that $k_A$ and $k_B$ never coexist at the same worldsheet time, so they cannot collide and make an intermediate state.\\
\indent These explanations are heuristic but we find them both amusing and useful, and we thought that the reader might as well. None of the main conclusions of this paper depend on the pictures we have drawn.

\subsection{A shortcut to the shortest time delay}\label{shortcut}
When we analyze the five point amplitude in the following sections, we will mainly be interested in the smallest phase shift. For this purpose it is sufficient to use standard techniques for approximating the string amplitudes in the Regge limit, which we will briefly review here (see e.g. \cite{bpst}). \\
\indent First consider the Virasoro-Shapiro amplitude,
\begin{align}
A(s,t)=\frac{g_{\text{c}}^2}{\alpha'}\int d^2z\, |z|^{-4-\alpha' t/2}|1-z|^{-4-\alpha's/2}.
\end{align}
This converges as long as $\text{Re }s,\text{Re }t, \text{Re }u<-4/\alpha'$, and is defined by analytic continuation elsewhere.    
If we take $\text{Im }s\to \infty$, then the integral is dominated by short worldsheet distances, at $z\sim 1/(\alpha' s)$. 
In this regime, the integral  
\begin{align}
A(s,t)&\sim g_{\text{c}}^2 \int d^2z\, |z|^{-4-\alpha' t/2}\exp(\alpha' s(z+\overline{z})/4).
\end{align}
This integral can be done using the integral representation for the gamma function, 
\be\label{gammaint}
\Gamma(\kappa) = \int_0^\infty dx \, x^{\kappa-1}e^{-x}
\ee
and one finds 
\begin{align}
A(s,t)\sim \frac{g_{\text{c}}^2}{\alpha'}\left(1-e^{-i\pi \alpha' t/2}\right)\Gamma(-1-\alpha' t/4)^2(\alpha' s/4)^{2+\alpha' t/2}.
\end{align}
This indeed captures the smallest phase shifts analyzed above.\\
\indent A similar procedure works to some extent for open strings. For instance, the integral for $A_{st}$ is 
\begin{align}
A_{st}(s,t)=\frac{g_{\text{o}}^2}{\alpha'}\int_{0}^{1} dy\, y^{-2-\alpha' t}(1-y)^{-2-\alpha' s},
\end{align}
which is dominated by $y\sim 1/(\alpha' s)$, again defining it by continuation from large imaginary $s$. Since the integral is exponentially suppressed away from $y=0$, we are free to extend the limits of integration to infinity. This gives
\begin{align}\label{openshort}
\frac{g_{\text{o}}^2}{\alpha'}\int_{0}^{\infty} dy\, y^{-2-\alpha' t}\exp(\alpha' sy)&=\frac{g_{\text{o}}^2}{\alpha'}e^{-i\pi \alpha' t}\Gamma(-1-\alpha' t)(\alpha' s)^{1+\alpha' t},
\end{align}
obtained as follows.  Given that we are working at large imaginary $s\sim e^{i\pi/2}|s|$, we rotate $y=i y_{\text{E}}$, giving a convergent integral in $y_\text{E}=x/(\alpha'|s|)=x/(\alpha' e^{-i\pi/2}s)$.  Matching this to the integral representation of the $\Gamma$ function (\ref{gammaint}) gives the right hand side of (\ref{openshort}), including the phase.  

This again matches the answer from the previous section.  However, we would like to caution the reader against using this shortcut indiscriminately. For instance, the integral for $A_{su}$ is not dominated by short worldsheet distances, and therefore the full amplitude must be analyzed.

 \subsection{Long strings in the $s$-channel}\label{longstrings}
Now that we have computed the time delays and peak impact parameters, it is natural to try to reproduce their precise coefficients from the $s$-channel picture. To do so, we need to find a on-shell long string which obeys the physical state conditions. Also, its energy and angular momentum must equal those of the initial state,
\begin{align}
E_{i}=2E,\hspace{10 mm}J_i=bE.
\end{align}
Although the squared mass of a state on a worldsheet is quantized in integer multiples of $1/\alpha'$, at large mass the spacing between the masses goes as $\Delta M=1/(\alpha' M)\to 0$, so there exist classical solutions with masses arbitrarily close to $2E$. This suggests that it might be possible to describe the scattering process by a single classical solution, although such a description is by no means guaranteed.\\
\indent Let us start with the case of backscattering at $b=0$, corresponding to the open string ordering $A_{su}$, as discussed in \cite{sst}. The angular momentum of the intermediate state vanishes, so we expect to make a long string that oscillates back and forth in the $x$-direction. Such a solution has been studied extensively in the literature \cite{yoyo1,yoyo2,yoyo3}, and is known as the yo-yo. It is simplest to describe this solution in static gauge $\tau=X^0$, where there is a well-known method for constructing on-shell solutions \cite{yoyo3}. Let $\vec{Y}(\tau)$ be the trajectory of one of the endpoints of the string, with $\vec{Y}(\tau+P)=\vec{Y}(\tau)$ for some period $P$. Then an on-shell solution is given by
\begin{align}
\vec{X}(\tau,\sigma)=\frac{1}{2}(\vec{Y}(\tau+\sigma)+\vec{Y}(\tau-\sigma)),\hspace{10 mm}0\le \sigma\le \frac{P}{2}.
\end{align}
For the yo-yo, we choose the trajectory 
\begin{align}
Y^1(\tau)=|L-\tau|,\hspace{10 mm}0\le \tau\le 2L.
\end{align}
This function can be extended to a periodic function of $\tau$, with period $P=2L$. The energy of the corresponding on-shell string must be equal to the energy of the initial state, so
\begin{align}
2E=\frac{1}{2\pi \alpha'}\int_0^{L} d\sigma\, \dot{X}^0=\frac{L}{2\pi \alpha'}.
\end{align}
The strings should be able to backscatter after half of an oscillation of the intermediate string, corresponding to a time delay $T=L$. This exactly reproduces the smallest time delay in the backscattering amplitude (\ref{timedelaysu}).\\
\indent The generalization to closed strings is straightforward; we simply glue two copies of the yo-yo together at their endpoints. The length of the intermediate state is then 
\begin{align}
L=2\pi \alpha' E.
\end{align}
We therefore expect time delays of the form $2\pi n \alpha' E$. This explains the numerical coefficient in (\ref{noangles}).\\
\begin{figure}
\begin{center}
 \begin{tikzpicture}[scale=1.5]
 \draw[>->,color=red] (-1.2,.5) -- (.31,.5) -- (1,-1);
  \draw[>->,color=blue] (1.2,-.5) -- (-.31,-.5) -- (-1,1);
  \draw (-.2,.7) node {$d$};
  \draw[dashed] (-.31,-.5) -- (-.31,.5);
    \draw[dashed] (-.31,-.5) -- (.59,-.08);
        \draw[dashed] (-.31,-.5) -- (-1,-.5);
 \draw (.3,-.33) node {$b$};
  \draw (-.23,0) node {$b$};
    \draw (-1.5,0) node {$(a)$};
    \draw (-1.1,.35) node {$k_A$};
        \draw (1.1,-.35) node {$k_B$};
        \draw (.8,-.9) node {$k_1$};
                \draw (-.8,.9) node {$k_3$};
    \draw (-.5,-.4) node {$\theta$};
   \end{tikzpicture}
\hspace{10 mm}
\begin{tikzpicture}[scale=1.5]
 \draw[>->,color=red] (-1,0) -- (0,0) -- (.46,-1);
  \draw[>->,color=blue] (1,0) -- (0,0) -- (-.46,1);
  \draw (.76,-.5) -- (-.76,.5);
      \draw (-1.5,0) node {$(b)$};
   \end{tikzpicture}
\hspace{10 mm}
 \begin{tikzpicture}[scale=1.5]
 \draw[>->,color=red] (-1.2,.5) -- (.8,.5) -- (.11,-1);
  \draw[>->,color=blue] (1.2,-.5) -- (-.8,-.5) -- (-.11,1);
    \draw (-1.5,0) node {$(c)$};
    \draw (-1.1,.35) node {$k_A$};
        \draw (1.1,-.35) node {$k_B$};
        \draw (.4,-.9) node {$k_1$};
                \draw (-.4,.9) node {$k_3$};
    \draw (-.6,-.35) node {$\theta$};
\end{tikzpicture}
   \end{center}
\caption{\label{traceback} $(a)$ The asymptotic trajectories of the incoming and outgoing states, traced back to the collision at $T=0$. The endpoints of the intermediate state follow the rhombus in the center of the diagram. $(b)$ The process is suppressed at $b=0$, since the endpoints of the intermediate state at a given snapshot of time do not hit the red and blue lines. $(c)$ The intermediate state corresponding to the backscattering ordering $A_{su}$.}
\end{figure}
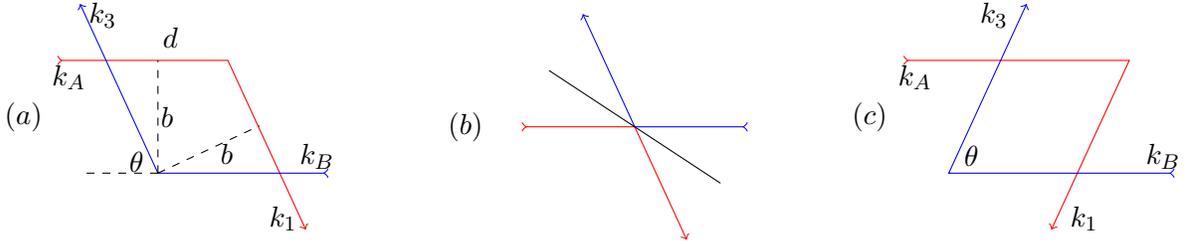
\indent Now let us try to generalize to nonzero impact parameter, starting with the smallest phase shift $e^{-i\pi \alpha' t}$ for open strings. To do this we need to choose the trajectories of the endpoints of the classical $s$-channel state. In Figure \ref{traceback}$a$ we have traced the asymptotic trajectories of the initial and final strings straight back to the place where they join. Conservation of angular momentum requires that the initial and final impact parameters are equal, so that these traced-back trajectories form a rhombus. Let us consider an $s$-channel state whose endpoints follow the edges of the rhombus. The oscillation period is then $4d$, so the energy of the state is 
\begin{align}
\frac{1}{2\pi \alpha'}\int_{0}^{2d} d\sigma=\frac{d}{\pi \alpha'}.
\end{align}
Matching to the initial state, we have $d=2\pi \alpha' E$. It follows from elementary geometry that $b=-2\pi \alpha' E\sin\theta$, which reproduces the answer (\ref{peak4ptsopen}). At any other impact parameter, the endpoints of the intermediate state that we have discussed do not hit the trajectories of the center of masses of the incoming strings, as shown in Figure \ref{traceback}$b$.\\
\indent Next we can compute the time delay of String 1, which we assume is released from the bottom right corner of the rhombus as shown in Figure \ref{traceback}$a$. The red path turns instantaneously at an $x$ position 
\begin{align}
x_0=-\frac{b(1-\cos\theta)}{2\sin\theta}.
\end{align}
The trajectory of String 1 is then
\begin{align}
x(T)&=(T-x_0)\cos\theta+x_0\\
y(T)&=(T-x_0)\sin\theta+b/2,
\end{align}
so its $x_1$ coordinate satisfies
\begin{align}
x_1(T)&=x(T)\cos\theta+y(T)\sin\theta=T+b\tan(\theta/2).
\end{align}
The time delay is therefore
\begin{align}
\Delta T=b\tan(\theta/2)=2\pi\alpha'  E(1-\cos\theta),
\end{align}
which is the correct answer (\ref{timedelayst}).\\
\indent The string whose endpoints trace out the rhombus cannot be the full description of the intermediate state. The angular momentum of this string is \cite{yoyo3}
\begin{align}
J=\frac{\text{Area swept in one revolution}}{4\pi \alpha'}=\frac{bE}{2},
\end{align}
which is half of the angular momentum of the initial state. One possibility is that the remaining spin is carried by oscillations on top of the long string. At oscillator level $n$, the largest spin is $n$, so oscillators at level $n=bE/2$ would need to be excited in order to account for the missing angular momentum. With this additional excitation, the energy of the intermediate state becomes
\begin{align}
\sqrt{(2E)^2+\alpha' n}\approx 2E\left(1+\frac{\alpha' b}{8E}\right).
\end{align}
At small angles the peak impact parameter satisfies $b\ll \alpha' E$, so the second term in the parentheses is negligible, and the intermediate state is approximately on-shell. Regardless, this solution seems to reproduce several nontrivial features of the $2\to2$ amplitude.\\  
\indent An analogous picture works for the backscattering diagram $A_{su}$ at finite scattering angle. In this case the string is made as shown in Figure \ref{traceback}$c$, with the rhombus reflected about the $x$-axis. 

\subsection{Putting it together:  S-matrix data and the scattering geometry for $A_{st}$}

Let us focus on the amplitude $A_{st}$.   The geometry indicated in Figure \ref{traceback}$a$ satisfies several overconstrained tests in the S-matrix `data' that we have developed in previous sections, as well as surviving an additional test at five points.    Let us pause to summarize this here.  

The peak trajectory 1 traces back to meet trajectory A (consistently in all directions), and similarly for trajectories 3 and B.  This is as expected from the Reggeon analysis in Section \ref{shortcut}.  There is a `yo yo' solution which is a good candidate for the produced $s$-channel state; as described in the previous section this fits nontrivially with the geometry in Figure \ref{traceback}$a$.        

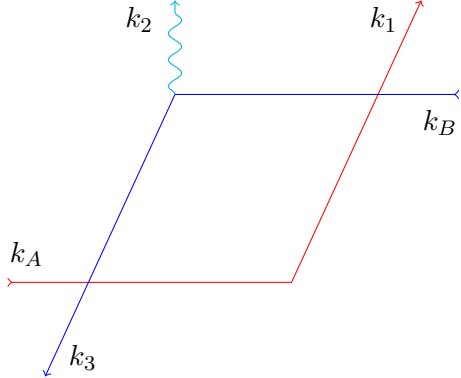
\begin{figure}
\begin{center}
\begin{tikzpicture}[scale=2.5]
 \draw[>->,color=red] (-1.2,-.5) -- (.31,-.5) -- (1,1);
  \draw[>->,color=blue] (1.2,.5) -- (-.31,.5) -- (-1,-1);
      \draw (-1.1,-.35) node {$k_A$};
        \draw (1.1,.35) node {$k_B$};
        \draw (.8,.9) node {$k_1$};
                \draw (-.8,-.9) node {$k_3$};
 \path [->,draw=cyan,snake it]
    (-.31,.5)  -- (-.31,1);
      \draw (-.5,.9) node{$k_2$};
\end{tikzpicture}
\end{center}
\caption{\label{5ptpics}  At five points, a low energy string 2 is emitted from String B at the point where it turns.}
\end{figure}

Finally, it turns out that this simple geometry also survives a nontrivial test involving Bremsstrahlung radiation at five points.  We will analyze this in detail in the next section, but for now simply note that addition of a radiation leg to the $A_{st}$ process at four points gives 
 the result depicted in Figure \ref{5ptpics}.
That is, the radiation emerges from the turning point of the trajectory of B into 3 (and similarly for A into 1).  

 In the next section, we will take this geometry as given and discuss the role of longitudinal versus transverse spreading in the process.   

\subsection{When is the string made? Longitudinal vs. transverse spreading}\label{longvstrans}

In the previous sections we have found a nonzero peak impact parameter in the $A_{st}$ diagram, and found a simple solution for the created intermediate strings. It is natural to ask how this fits in with the Gaussian-distributed transverse spreading of strings calculated in
\cite{lennyspreading}.    As we will review momentarily, the calculation of \cite{lennyspreading} leads to a Gaussian density for the distance $x_\perp$ between the endpoint of an open string and the center of mass of the string,
\begin{align}\label{standardtransverse}
\rho(x_\perp)=\exp\left(-\frac{x_{\perp}^2}{2\alpha' \log \frac{s}{s_0}}\right),
\end{align}
for constant $s_0$.  Creation of the intermediate string on the tail of this transverse distribution would not be consistent with the observed nonzero value of the peak impact parameter, since the transverse spreading is suppressed by the Gaussian.   

We will elaborate on this shortly as well as further in Appendix \ref{transverseappendix}.   
First let us briefly review the argument for (\ref{standardtransverse}). For closed strings, the variance of the transverse embedding coordinate in the free string ground state is 
\begin{align}\label{sumtext}
\langle (\Delta X_\perp)^2\rangle=\alpha' \sum_{n=1}^{n_{\text{max}}}\frac{1}{n}=\alpha'\log \frac{n_{\text{max}}}{n_0}+ O(1/n_{\text{max}}) ,
\end{align}
where $n_{\text{max}}$ is the highest frequency that is probed by the measurement and $n_0$ is a constant\footnote{which is given explicitly by the calculation in Appendix \ref{transverseappendix}, equations (\ref{transversesum})-(\ref{sumapprox}).}.  For open strings, this holds for the endpoints of the strings (see \cite{usBH}\ for a detailed review).  

To give this computation an operational meaning in the context of string scattering, one can compute the Veneziano amplitude in the Regge limit, in the transverse brick wall frame $p_A^+=-p_1^+$ and $p_B^+=-p_3^+$. As shown in \cite{bpst}, in this frame the worldsheet length of each of the strings is conserved by the interaction, and the amplitude reduces to the computation of simple quantum mechanical expectation values in the free string ground state. The interaction provides a cutoff on the mode sums at 
\begin{align}\label{nmax}
n_{\text{max}}\sim -\frac{s}{t},
\end{align}
which is manifest in Equation (4.20) of \cite{bpst}\  (for the regime $s\gg -t\gg 1/\alpha'$).  As a result, we obtain the correct amplitude if we cut off the infinite mode sum in the incoming string wavefunction at $n_{\text{max}}$.    
Keeping track of the transverse string spreading $x_\perp$ between one of the endpoints and the center of mass, we obtain a probability distribution $\rho(x_\perp)=\text{Tr}|\Psi(x_\perp,\{ {\hat X_I}\})|^2$ where the trace is over the other degrees of freedom $\{\hat X_I\}$.  This distribution $\rho(x_\perp)$ is given by (\ref{standardtransverse}).  In Appendix \ref{transverseappendix}\ we elaborate on the structure of this Gaussian distribution and its relation to the explicit mode sum (\ref{sumtext}), in the context of alternative proposals for transverse string spreading which agree at leading order at large $n_{\text{max}}$, but disagree with the structure of the subleading corrections.  

\begin{figure}
\begin{center}
\begin{tikzpicture}[scale=1.5]
\draw (-1,1) -- (1,1);
\draw (-1,.75) node {$k^+_A$};
\draw (-1,.25) node {$k^+_B$};
\draw (1,.75) node {$k^+_1$};
\draw (1,.25) node {$k^+_3$};
\draw (-1,.5) -- (-.1,.5);
\draw (.1,.5) -- (1,.5);
\draw (-1,0) -- (1,0);
\end{tikzpicture}
\end{center}
\caption{The worldsheet diagram in light-cone gauge for the ordering $A_{st}$ in the transverse brick wall frame, where the length of each of the strings is conserved. The interaction occurs during a short time in the Regge limit.}
\end{figure}
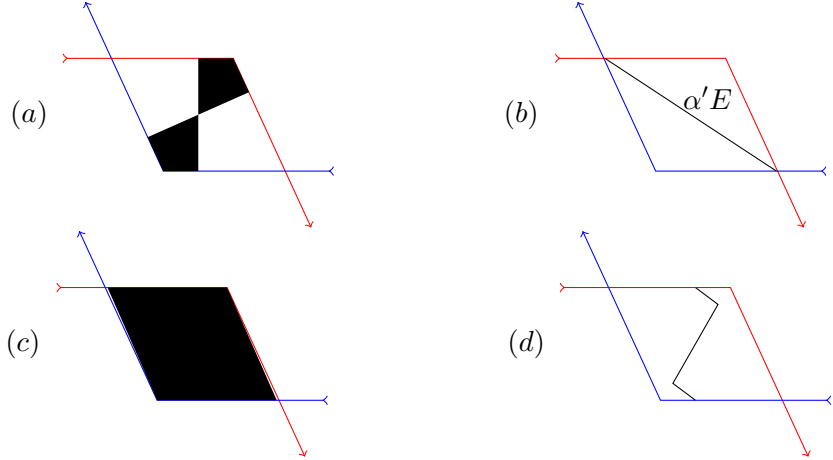
\begin{figure}
\begin{center}
\begin{tikzpicture}[scale=1.5]
 \draw[>->,color=red] (-1.2,.5) -- (.31,.5) -- (1,-1);
  \draw[>->,color=blue] (1.2,-.5) -- (-.31,-.5) -- (-1,1);
\draw (-1.5,0) node {$(a)$};
  \fill (0,.5) -- (.31,.5) -- (.45,.2) -- (0,0);
\fill (0,-.5) -- (-.31,-.5) -- (-.45,-.2) -- (0,0);
\end{tikzpicture}\hspace{20 mm}
 \begin{tikzpicture}[scale=1.5]
 \draw[>->,color=red] (-1.2,.5) -- (.31,.5) -- (1,-1);
  \draw[>->,color=blue] (1.2,-.5) -- (-.31,-.5) -- (-1,1);
  \draw (.76,-.5) -- (-.76,.5);
      \draw (-1.5,0) node {$(b)$};
  \draw (.15,.15) node {$\alpha' E$};
   \end{tikzpicture}\\
\begin{tikzpicture}[scale=1.5]
 \draw[>->,color=red] (-1.2,.5) -- (.31,.5) -- (1,-1);
  \draw[>->,color=blue] (1.2,-.5) -- (-.31,-.5) -- (-1,1);
\draw (-1.5,0) node {$(c)$};
  \fill (-.75,.5) -- (.31,.5) -- (.75,-.5) -- (-.31,-.5);
\end{tikzpicture}
\hspace{20 mm}
\begin{tikzpicture}[scale=1.5]
 \draw[>->,color=red] (-1.2,.5) -- (.31,.5) -- (1,-1);
  \draw[>->,color=blue] (1.2,-.5) -- (-.31,-.5) -- (-1,1);
\draw (-1.5,0) node {$(d)$};
\draw (0,.5) -- (.2,.35) -- (-.2,-.35) -- (0,-.5);
\end{tikzpicture}
\caption{\label{transversecreation}$(a)$ If the intermediate state is created and destroyed when the initial and final states are purely transverse separated, then the intermediate string traces out the region shown in black. $(b)$ Longitudinal spreading allows production of long strings when the initial states are separated by a distance $\sim \alpha' E$. $(c)$ The filled in region is traced out by the intermediate state if the string is made as in $b$. $(d)$ A snapshot of the classical string discussed in Section \ref{longstrings} at the time when the endpoints are at the same $x$ position.}
\end{center}
\end{figure}

\indent Now let us return to the four-point ordering $A_{st}$, where strings are created at finite impact parameter. One simple possibility is that the intermediate state is created when the incoming states are purely transverse separated with respect to the $x$ direction of their relative motion, and then breaks when the final states are transverse separated with respect to the $x_1$ direction of the outgoing strings' relative motion. This scenario is depicted in Figure \ref{transversecreation}$a$.  A more extreme possibility would be that the string is created and then immediately decays at the shorter diagonal of the rhombus.  This corresponds to when it is transversely extended with respect to the brick wall frame (\ref{brickwall}). \\
\indent Assuming the transverse spreading formula (\ref{standardtransverse}), we can give an indirect test of Figure \ref{transversecreation}$a$. The only way a string can be created is if the endpoints of the incoming strings join. In Appendix \ref{transverseappendix} we will discuss the possibility that the strings cannot be treated as classical spacetime source distributions, but let us first assume that this classical picture is correct. 
Then the probability for the endpoints to join at impact parameter $b$ is 
\begin{align}\label{transversesuppression}
\int dx_\perp\, \rho(x_\perp)\rho(b-x_\perp)\propto \exp\left(-\frac{b^2}{ 2\alpha'\log s}\right).
\end{align} 
If Figure \ref{transversecreation}$a$ is an accurate representation of the process, then we should see a corresponding suppression factor in the scattering amplitude convolved with localized wavepackets. Recall that the ordering $A_{st}$ is peaked at $b=-2\pi \alpha' E\sin\theta$.  As we ramp up the scattering angle, the amplitude should contain a suppression factor of the form (\ref{transversesuppression}), with $b=-2\pi \alpha' E\sin\theta$.  For $s\gg -t\gg 1/\alpha'$, one finds
\begin{align}\label{amplst}
A_{st}=(-t)^{-3/2}e^{\alpha' t}e^{-i\pi \alpha' t}\left(-\frac{s}{t}\right)^{\alpha' t}(1+O(t/s,1/(\alpha' t))),
\end{align}
where we are focusing on the first oscillation in the expansion of (\ref{fullforms}). The magnitude of the scattering amplitude of wavepackets localized at $b=-2\pi \alpha' E\sin\theta$ is equal to (\ref{amplst}), and does not contain the expected suppression factor.


Therefore it seems that either the standard transverse spreading formula (\ref{standardtransverse}) is incorrect, or that Figure \ref{transversecreation}$a$ -- with the string joining by virtue of transverse spreading -- is not an accurate picture.   Similarly, an instantaneous joining and splitting at zero transverse separation in the brick wall frame would seem to be contraindicated.  Given the standard arguments for the logarithmic growth of strings, let us assume that (\ref{standardtransverse}) is correct, and try to look for another picture of the process. One possibility is that strings also have a longitudinal size, as advocated in \cite{lennyspreading}\ based on the analogous calculation to (\ref{sumtext}) for the longitudinal embedding coordinates, the second equation in (\ref{spreadapprox}).  To evaluate the corresponding suppression factor in the amplitude in a similar way, one would need the distribution for the longitudinal direction $X^+$ analogous to the Gaussian (\ref{standardtransverse}) given above for $X_\perp$; more generally the joint distribution $\rho(X^+, X_\perp)$ would enter.    

The constraint equation relating the longitudinal and transverse directions is linear in $X^+$ and quadratic in $X_\perp$, while the ground state wavefunction is Gaussian in $X_\perp$.  We might therefore expect that at large $X^+$, $\rho(X^+)\approx\exp(-|X^+|/\sqrt{\langle (X^+)^2\rangle})$.  In this case, the suppression factor analogous to (\ref{transversesuppression}) for an interaction at $X^+\sim \alpha' E$ in the center of mass frame would be of order $ e^{\alpha' t}$, agreeing with
the corresponding factor in (\ref{amplst}).


In other words, given what is known about the longitudinal distribution, it may be possible to create a string with length $\sim \alpha' E$ along the long axis of the rhombus, as shown in Figure \ref{transversecreation}$b$. 
Longitudinal spreading therefore provides a consistent picture of the scattering process at finite impact parameter. As a possible further check of this picture, recall that the integral over the light-cone time $\Delta X^-$ between the joining and splitting interactions is dominated at 
\begin{align}
\Delta X^-\sim -\frac{\alpha' t}{p_A^+}.
\end{align}
when the amplitude is defined via appropriate analytic continuation \cite{bpst,usBH}.
Although this is a short light-cone time, String A travels a long distance in the light-cone space direction during this time \cite{usBH},
\begin{align}
\Delta X^+\sim-\Delta X^- \frac{(p_A^+)^2}{t}\sim \alpha' p_A^+,
\end{align}
which is of order $\alpha' E$ in the center of mass frame. This is consistent with the idea that the intermediate state is created and destroyed at a light cone distance $\Delta X^+\sim \pm \alpha' E$ in the center of mass frame.   Since this saddle point requires analytic continuation, it does not directly describe the real time process, but it is interesting that the complex time and distance scales that come in line up with those in our real time picture of the process.  Similar comments apply to the results in \cite{grossmende}.  We do not claim to have given a derivation of longitudinal spreading at four points, but it seems that some effect beyond the usual transverse spreading (\ref{standardtransverse}) is needed for consistency of the amplitude. \\
\indent In fact, if we assume that the intermediate state is the classical long string discussed in Section \ref{longstrings}, it is straightforward to exclude an interaction that is purely transverse (as defined by the direction $x$ of relative motion of the incoming strings). When the endpoints of the intermediate string are at the same $x$-position, the long string is not extended only in a transverse direction. Instead, it has two kinks with longitudinal extent, as shown in Figure \ref{transversecreation}$a$. Therefore the incoming strings could not join at $T=0$ to form the intermediate state if they only had transverse size. On the other hand, Figure \ref{transversecreation}$b$ is an accurate snapshot of the incoming state at a fixed time (the string also has some transverse momentum distributed throughout its length at this time).  This argument alone, however, does
not exclude an instantaneous joining and splitting at transverse separation as defined by the brick wall frame.   

\indent We note that the Fourier transform of the amplitude with respect to $t$ does exhibit a transverse suppression factor $\exp(-b^2/(\alpha'(\log s-i\pi)))$.   However, this Fourier transform is different from the scattering amplitude of localized wavepackets at angle $\theta$, and in itself it does not admit an immediate interpretation in terms of the distribution of string.\footnote{We thank Steve Giddings for extensive discussions of this point.}  
In Appendix \ref{transverseappendix} we elaborate on this,  also discussing an alternative possibility for a cut off string wavefunction motivated by the Fourier transform of the amplitude.  This alternative form combined with a simple joining interaction would generate the required transverse effect, peaked at the correct impact parameter.   But as we explain in Appendix \ref{transverseappendix}, it is distinct from the Gaussian wavefunction of width $\sqrt{\alpha'\log n_{\text{max}}}$ which arises from the transverse mode sum \cite{lennyspreading}\ cut off at $n_{\text{max}}$.  We show there by explicit calculation that the two differ at the first subleading correction at large $n_{\text{max}}$, with the Gaussian wavefunction of width $\sqrt{\alpha'\log n_{\text{max}} }$ being the one which arises from the transverse mode sum simply cut off at $n_{\text{max}}$.  The latter statement, in turn, has substantial support from the analysis in \cite{bpst}, as mentioned above.  We therefore find it very plausible that the transverse distribution (\ref{standardtransverse}) is correct, although we will continue to present the results as conditioned on this assumption.    

\section{Five-point function:  open string radiation}\label{fiverad}

We will now generalize our analysis of Regge amplitudes and their phases to the case of five-point diagrams with an additional outgoing leg.
In the presence of this outgoing radiation, the time shifts for the strings in the underlying four-point amplitude will adjust, depending on the energy $E_2$ and angle $\theta_2$ of the radiation.  
In particular, in the ordering $A_{tu}$ described above, there were no time shifts at the four-point level, and it is interesting to ask which direction the time shifts go in the presence of the radiation.    

We will find examples of both delays and advances in the five-point diagrams whose topologies correspond to this four-point amplitude.   The advance is particularly important for our assessment of longitudinal nonlocality, and we will describe in detail how it follows from several concrete features of the amplitude, combined
with two assumptions which we find plausible (and follow from standard calculations, e.g. in \cite{bpst}).  In particular, the peak trajectories derived from a wavepacket analysis will reveal a regime of kinematics in which the outgoing trajectory of String 1, traced back in time,  emerges from the origin before the putative $T=0$ collision.  Meanwhile, the dominant contribution to the worldsheet vertex operator integral comes from String 1 emerging from A after the latter emits a Reggeon.
Taking from this last piece of `data' that the trajectory of 1 is a continuation of that of A, and taking as given the standard $\alpha' \log s$ range of transverse spreading, we will show that an early interaction follows by causality.

In order to assess the time shifts at five points, we will compute the bosonic string amplitude for five tachyons, in a kinematic regime where they are massless to good approximation. This requires that the energies of all the strings be much greater than $1/\sqrt{\alpha'}$.  For the question about early radiation, we can consider the regime
\be\label{Etwo}
\frac{1}{\sqrt{\alpha'}}\ll E_2\ll E.
\ee  
That is, the strings $\text{A}, \text{B}, 1,$ and $3$ have energy of order $E$, whereas String $2$ has the much smaller energy $E_2$.  In this sense, we may think about String 2 as a perturbation -- extra outgoing radiation -- on top of the underlying $2\to2$ scattering amplitude $\text{AB}\to 13$.   In the center of mass frame, whose kinematics is described in Appendix \ref{kinematics},  we will work in the Regge regime with $E\theta_1$ fixed, ensuring that the amplitude is not strongly suppressed as is the case at hard scattering. This kinematical regime is known as the double-Regge limit in the literature, for reasons that will become clear momentarily.


The five-point amplitude we require was proposed by Bardakci and Ruegg in the context of dual models, before string theory was developed \cite{BR}.   We will make use of the analysis of Bialas and Pokorski in \cite{BP}\ after first reproducing the Bardakci-Ruegg amplitude in perturbative open string theory.     
\subsection{Shortcut analysis and double Regge regime}\label{naive}
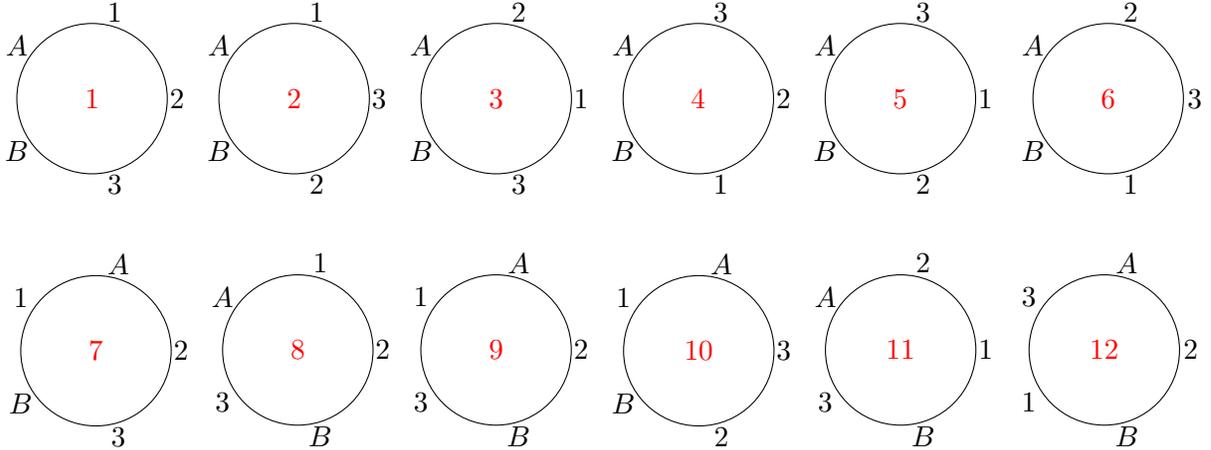
\begin{figure}
\begin{center}
 \begin{tikzpicture}
\draw (0,0) circle (1 cm);
\draw[color=red] (0,0) node{$1$}; 
\draw (1.13,0) node {$2$};
\draw (.3,1.15) node {$1$};
\draw (-1,.7) node {$A$};
\draw (-1,-.7) node {$B$};
\draw (.3,-1.15) node {$3$};
\end{tikzpicture}
\hspace{-2 mm}
 \begin{tikzpicture}
\draw (0,0) circle (1 cm);
\draw[color=red] (0,0) node{$2$}; 
\draw (1.13,0) node {$3$};
\draw (.3,1.15) node {$1$};
\draw (-1,.7) node {$A$};
\draw (-1,-.7) node {$B$};
\draw (.3,-1.15) node {$2$};
\end{tikzpicture}
\hspace{-2 mm}
 \begin{tikzpicture}
\draw (0,0) circle (1 cm);
\draw[color=red] (0,0) node{$3$}; 
\draw (1.13,0) node {$1$};
\draw (.3,1.15) node {$2$};
\draw (-1,.7) node {$A$};
\draw (-1,-.7) node {$B$};
\draw (.3,-1.15) node {$3$};
\end{tikzpicture}
\hspace{-2 mm}
 \begin{tikzpicture}
\draw (0,0) circle (1 cm);
\draw[color=red] (0,0) node{$4$}; 
\draw (1.13,0) node {$2$};
\draw (.3,1.15) node {$3$};
\draw (-1,.7) node {$A$};
\draw (-1,-.7) node {$B$};
\draw (.3,-1.15) node {$1$};
\end{tikzpicture}
\hspace{-2 mm}
 \begin{tikzpicture}
\draw (0,0) circle (1 cm);
\draw[color=red] (0,0) node{$5$}; 
\draw (1.13,0) node {$1$};
\draw (.3,1.15) node {$3$};
\draw (-1,.7) node {$A$};
\draw (-1,-.7) node {$B$};
\draw (.3,-1.15) node {$2$};
\end{tikzpicture}
\hspace{-2 mm}\nolinebreak
 \begin{tikzpicture}
 \nolinebreak
\draw (0,0) circle (1 cm);
\draw[color=red] (0,0) node{$6$}; 
\draw (1.15,0) node {$3$};
\draw (.3,1.15) node {$2$};
\draw (-1,.7) node {$A$};
\draw (-1,-.7) node {$B$};
\draw (.3,-1.15) node {$1$};
\end{tikzpicture}\\\vspace{5 mm}
 \begin{tikzpicture}
\draw (0,0) circle (1 cm);
\draw[color=red] (0,0) node{$7$}; 
\draw (1.13,0) node {$2$};
\draw (.3,1.15) node {$A$};
\draw (-1,.7) node {$1$};
\draw (-1,-.7) node {$B$};
\draw (.3,-1.15) node {$3$};
\end{tikzpicture}
\hspace{-2 mm}
 \begin{tikzpicture}
\draw (0,0) circle (1 cm);
\draw[color=red] (0,0) node{$8$}; 
\draw (1.13,0) node {$2$};
\draw (.3,1.15) node {$1$};
\draw (-1,.7) node {$A$};
\draw (-1,-.7) node {$3$};
\draw (.3,-1.15) node {$B$};
\end{tikzpicture}
\hspace{-2 mm}
 \begin{tikzpicture}
\draw (0,0) circle (1 cm);
\draw[color=red] (0,0) node{$9$}; 
\draw (1.13,0) node {$2$};
\draw (.3,1.15) node {$A$};
\draw (-1,.7) node {$1$};
\draw (-1,-.7) node {$3$};
\draw (.3,-1.15) node {$B$};
\end{tikzpicture}
\hspace{-2 mm}
 \begin{tikzpicture}
\draw (0,0) circle (1 cm);
\draw[color=red] (0,0) node{$10$}; 
\draw (1.13,0) node {$3$};
\draw (.3,1.15) node {$A$};
\draw (-1,.7) node {$1$};
\draw (-1,-.7) node {$B$};
\draw (.3,-1.15) node {$2$};
\end{tikzpicture}
\hspace{-2 mm}
 \begin{tikzpicture}
\draw (0,0) circle (1 cm);
\draw[color=red] (0,0) node{$11$}; 
\draw (1.13,0) node {$1$};
\draw (.3,1.15) node {$2$};
\draw (-1,.7) node {$A$};
\draw (-1,-.7) node {$3$};
\draw (.3,-1.15) node {$B$};
\end{tikzpicture}
\hspace{-2 mm}\nolinebreak
 \begin{tikzpicture}
 \nolinebreak
\draw (0,0) circle (1 cm);
\draw[color=red] (0,0) node{$12$}; 
\draw (1.15,0) node {$2$};
\draw (.3,1.15) node {$A$};
\draw (-1,.7) node {$3$};
\draw (-1,-.7) node {$1$};
\draw (.3,-1.15) node {$B$};
\end{tikzpicture}
\end{center}
\caption{\label{orderings}
The twelve open string orderings at five points.}
\end{figure}
In the open string theory tree-level five-point tachyon amplitude, there are 12 diagrams differing by the ordering of the vertex operators on the disk. These are depicted in \cite{BP}, which we reproduce here for convenience in Figure \ref{orderings}.  We will begin by computing Diagram 7; diagrams 1, 8, and 9 may be obtained from this by switching A with 1 and B with $3$.\footnote{The other eight orderings will not be important for us, as we will explain below.} In the next section we will compute this amplitude rigorously without approximation, but first let us use the shortcut of Section \ref{shortcut} to build some intuition and to determine the leading contribution to the diagram.  This leading contribution will play a role in our interpretation of the time shifts later in the paper.  

Define the kinematic invariants
\be\label{Knotation}
K_{IJ}\equiv 2\alpha' k_I\cdot k_J.
\ee
The behavior of these quantities in the center of mass frame in our kinematic regime is given in Appendix \ref{kinematics}.

According to the standard rules of string perturbation theory \cite{JoeBook}, 
\be\label{sevenint}
A_7=\frac{g_{\text{o}}^3}{\alpha'}\int_0^1 dy_2\int_0^{y_2} dy_A\,  y_A^{K_{A1}} y_2^{K_{12}} (1-y_A)^{K_{A3}}(1-y_2)^{K_{23}}(y_2-y_A)^{K_{A2}}.
\ee
where the $y_I$ are the positions of vertex operators on the boundary of the worldsheet; we have fixed $y_1=0, y_3=1,$ and $y_B=\infty$, leaving us to integrate over the positions of A and $2$.  

The first step is to make the change of variables $x=y_A/y_2$ to put the integrals on a more symmetric footing. This gives
\be
\int_0^1 dy_2\int_0^1 dx\,  x^{K_{A1}} y_2^{K_{B3}} (1-xy_2)^{K_{A3}}(1-y_2)^{K_{23}}(1-x)^{K_{A2}},
\ee
where we have used the identity 
\begin{align}\label{mandelidentity}
K_{A1}+K_{A2}+K_{12}=K_{B3}-1.
\end{align}
The integrals then converge when the exponents satisfy $\text{Re }K_{IJ}>-1$. The amplitude is defined elsewhere by analytic continuation from the domain of convergence.\\
\indent Following the shortcut from Section \ref{shortcut}, we take the imaginary parts of the large variables $\text{Im }K_{A3}, \text{Im }K_{A2},\text{Im }K_{23}$
to negative infinity in the ratio $\text{Im } K_{A3}\sim -E^2\to\infty, \text{Im } K_{A2}\sim -E\sim \text{Im }K_{23}$,  while holding the variables $K_{A1}$ and $K_{B3}$ fixed. The integral is then dominated by $x,y_2\sim 1/E$, so it may be approximated as 
\begin{align}\label{approx}
&\int_0^\infty dy_2\int_0^\infty dx\,  x^{K_{A1}} y_2^{K_{B3}} \exp(-K_{23}y_2-K_{A2}x-K_{A3}xy_2)\notag\\
&=K_{A3}^{-1-K_{A1}}K_{23}^{K_{A1}-K_{B3}} \Gamma(1+K_{A1})\Gamma(K_{B3}-K_{A1}){_1F_1}(1+K_{A1},1+K_{A1}-K_{B3},\kappa)\notag\\
&\hspace{5 mm}+K_{A3}^{-1-K_{B3}}K_{A2}^{K_{B3}-K_{A1}} \Gamma(1+K_{B3})\Gamma(K_{A1}-K_{B3}){_1F_1}(1+K_{B3},1+K_{B3}-K_{A1},\kappa).
\end{align}
Here $\kappa$ is defined as
\be\label{kappa}
\kappa = \frac{K_{23} K_{A2}}{K_{A3}} \sim -\alpha' E_2^2 \sin^2\theta_2.
\ee
and the phase is given by $K_{23}=e^{-i\pi}|K_{23}|$.  This last statement can be seen directly from the evaluation of the integral analogously to the discussion below (\ref{openshort}) above.   As at four points, we will find that this precisely matches the result for the same limit of the full amplitude, with the standard $i\epsilon$ prescription $K_{23}\to K_{23}-i\epsilon$, or similarly the inclusion of a decay width for the strings in the 2-3 $s$-channel.  

The origin of the term double-Regge limit is now clear. The dependence of (\ref{approx}) on the large kinematic invariants is of the form $s_1^{\alpha' t_1}s_2^{\alpha' t_2}$, where $s_1$ and $s_2$ are large and $t_1$ and $t_2$ are fixed.   This looks like the product of two single-Regge propagators. The amplitude has an additional functional dependence on the fixed $t$-like variables, which \cite{BP}\ interpret as a momentum-dependent vertex function.    
The fact that the dominant contribution to this integral has $x=y_A/y_2\to 0$ means that it is dominated by the regime where the vertex operator of String A is close to that of String 1, since $y_1=0$ and $y_2\le 1$. We therefore expect the trajectory of String A to be the continuation of the trajectory of String 1, in the sense that the two trajectories trace back to meet at some spacetime point. We will make use of this observation later in our analysis, in intepreting the peak time shifts derived from the phase of this amplitude.  

\subsection{Full analysis}
Now that we know what to expect, let us compute the amplitude without making any approximations within the integral, closely following the approach of \cite{BP}.  This will reproduce (\ref{approx}).   The integral (\ref{sevenint}) can be done explicitly in terms of generalized hypergeometric functions,
\begin{align}\label{fullformseven}
&B(1+K_{A2},1+K_{A1})B(1+K_{23},1+K_{B3})\notag \\
&\hspace{10 mm}\times {_3F_2}(-K_{A3}, 1+K_{A1}, 1+K_{B3}; 2+K_{A1}+K_{A2}, 2+K_{23}+K_{B3};1).
\end{align}
where
\be\label{Bdef}
B(a,b) = \frac{\Gamma(a)\Gamma(b)}{\Gamma(a+b)}.
\ee
The next step is to take the $_3F_2$ hypergeometric function and apply the transformation 
\begin{align}
&B(d-b,b)B(e-c,c)_3F_2(a,b,c;d,e;1)\notag\\
&=B(1-a,b)B(e-c,c-b)_3F_2(b,1+b-d,1+b-e;1+b-c,1+b-a;1)\notag\\
&\hspace{3 mm}+B(1-a,c)B(d-b,b-c)_3F_2(c,1+c-e,1+c-d;1+c-b,1+c-a;1).
\end{align}
Doing this and simplifying using a permutation of the identity (\ref{mandelidentity}) gives
\begin{align}\label{twoFs}
&B(1+K_{A3},1+K_{A1})B(1+K_{23},K_{B3}-K_{A1})\notag\\
&\hspace{5 mm}\times {_3F_2}(1+K_{A1}, -K_{A2}, 1+K_{B2}; 1+K_{A1}-K_{B3}, 2+K_{A1}+K_{A3}; 1) \notag\\
 &+B(1+K_{A3},1+K_{B3})B(1+K_{A2},K_{A1}-K_{B3})\notag\\
 &\hspace{5 mm}\times{_3F_2}(1+K_{B3}, -K_{23},1+K_{12}; 1+K_{B3}-K_{A1}, 2+K_{B3}+K_{A3};1).
\end{align}

The amplitude can now readily be approximated in double-Regge kinematics, using the limit
\be\label{Fthreelimit}
{_3F_2}(a, b,c; d, e; z)\to {_1F_1}\left(a, c; \frac{bc z}{e}\right) ~~ {\rm as} ~{b,c,e}\to \infty
\ee
to reduce the hypergeometric ${_3F_2}$ functions to hypergeometric ${_1F_1}$ functions. In particular, from Appendix \ref{kinematics} we have $|b|,|c| \sim EE_2$ and $e\sim E^2$ for the ${_3F_2}$ functions appearing in (\ref{twoFs}), along with $z=1$.    
Implementing this reduces (\ref{twoFs}) to 
\begin{align}\label{twoFones}
&B(1+K_{A3},1+K_{A1})B(1+K_{23},K_{B3}-K_{A1}){_1F_1}(1+K_{A1}, 1+K_{A1}-K_{B3},\kappa) \notag\\
 &+B(1+K_{A3},1+K_{B3})B(1+K_{A2},K_{A1}-K_{B3}){_1F_1}(1+K_{B3}, 1+K_{B3}-K_{A1},\kappa).
\end{align}
\indent At this point the only dependence on the large momenta is in the beta functions. As at four points, we can use Stirling's approximation for large $K_{A3}, K_{A2}$, and $K_{23}$ to give
\begin{align}\label{StirlingApprox}
B(1+K_{A3},1+K_{A1})&\sim \Gamma(1+K_{A1})(K_{A3})^{-1-K_{A1}}\nonumber\\
B(1+K_{A3},1+K_{B3})&\sim \Gamma(1+K_{B3})(K_{A3})^{-1-K_{B3}}\notag\\
B(1+K_{A2},K_{A1}-K_{B3})&\sim \Gamma(K_{A1}-K_{B3})(K_{A2})^{K_{B3}-K_{A1}}\notag\\
B(1+K_{23},K_{B3}-K_{A1})&\sim \Gamma(K_{B3}-K_{A1})(e^{-i\pi}|K_{23}|)^{K_{A1}-K_{B3}}.
\end{align}
The arguments of the beta functions in the first three lines are positive, enabling straightforward application of Stirling's approximation.
In the last line the argument is negative, and we derived the appropriate phase using the methods of Section \ref{fourpoints}.
Let us explain this explicitly here for completeness.  The negative kinematic variable $K_{23}\sim -\alpha' s_{23}$ is shifted at one loop to $K_{23}-i \alpha' m\Gamma$ as a result of decays of the intermediate state of mass $m$ that splits into strings 2 and 3.  Applying Euler's reflection formula to the beta function on the last line of (\ref{StirlingApprox}) gives
\be\label{spellout}
\hspace{-1 mm}B(1+K_{23},K_{B3}-K_{A1})=\frac{\sin(\pi (K_{B3}+K_{23}-K_{A1}))}{\sin(\pi K_{23})}\frac{\Gamma(K_{B3}-K_{A1})\Gamma(K_{A1}-K_{23}-K_{B3})}{\Gamma(-K_{23})},
\ee
with the arguments of the gamma functions now positive.  For the two gamma functions with large argument, we can apply Stirling's approximation and these produce a net factor $|K_{23}|^{K_{A1}-K_{B3}}$. Taking into account the nonzero decay rate, the leading behavior of the sine functions is
\be\label{reallyspellout}
\frac{e^{\pi\alpha'  m\Gamma}e^{i\pi(K_{23}+K_{B3}-K_{A1})}+\dots}{e^{\pi\alpha'  m\Gamma}e^{i\pi K_{23}}+\dots} \sim e^{-i\pi (K_{A1}-K_{B3})},
\ee 
where the terms $\dots$ on the top and bottom scale like $e^{-\pi\alpha'  m\Gamma}$. 
Focusing on the first term in the expansion, this gives the phase in the last line of (\ref{StirlingApprox}).  

 

Given this, we see that (\ref{twoFones}) reduces to the formula (\ref{approx}) derived in the previous section.
\subsection{Canceling the spurious poles}

Let us now take a closer look at the final result (\ref{approx}). This amplitude has a confusing feature: each of the two terms has poles at an infinite sequence of integer values of $K_{A1}-K_{B3}$.  These poles do not correspond to on-shell intermediate strings, so they must cancel between the two terms.  That they do cancel is most easily seen by expressing (\ref{approx}) in the form
\begin{align}\label{rearrange}
&\Gamma(1+K_{A1})\Gamma(1+K_{B3})K_{A3}^{-1-K_{A1}}K_{23}^{K_{A1}-K_{B3}} \left(\frac{\Gamma(K_{B3}-K_{A1})}{\Gamma(1+K_{B3})}{_1F_1}(1+K_{A1},1+K_{A1}-K_{B3},\kappa)\notag\right.\\
&\left.\hspace{31 mm}+(\kappa-i\epsilon)^{K_{B3}-K_{A1}}\frac{\Gamma(K_{A1}-K_{B3})}{\Gamma(1+K_{A1})}{_1F_1}(1+K_{B3},1+K_{B3}-K_{A1},\kappa)\right).
\end{align}
We have introduced the $i\epsilon$ prescription $\kappa-i\epsilon$, which follows from 
\begin{align}
\frac{K_{A2}^{K_{B3}-K_{A1}}(e^{-i\pi}|K_{23}|)^{K_{B3}-K_{A1}}}{K_{A3}^{K_{B3}-K_{A1}}}=(\kappa-i\epsilon)^{K_{B3}-K_{A1}}.
\end{align}

The sum of the terms in parantheses can be expressed in terms of the Tricomi hypergeometric function $U$, which is defined as
\be\label{UtwoFs}
U(a,b,z)=\frac{\Gamma(b-1)}{\Gamma(a)} z^{1-b}{_1F_1}(a-b+1,2-b,z)+\frac{\Gamma(1-b)}{\Gamma(a-b+1)}{_1F_1}(a,b,z).
\ee 
Using this relation, we finally obtain for Diagram 7 the result
\begin{align}\label{Seven}
A_7&= \frac{g_\text{o}^3}{\alpha'}\Gamma(1+K_{A1})\Gamma(1+K_{B3})K_{23}^{K_{A1}-K_{B3}}K_{A3}^{-1-K_{A1}}U(1+K_{A1},1+K_{A1}-K_{B3}, \kappa-i\epsilon),
\end{align}
with the phase of $K_{23}<0$ being $e^{-i\pi}$ as explained above. The tradeoff for the cancellation of the spurious poles is that $U$ has a more complicated analytic structure than $_1F_1$, as can be seen from the branch cut in $\kappa$ in (\ref{rearrange}). In the next section we will discuss a limit in which the analytic structure of $U$ simplifies, and the amplitude acquires a definite phase.

The calculations of Diagrams 1 and 8 proceed analogously, and we will just cite the results,
\begin{align}
A_1&=\frac{g_\text{o}^3}{\alpha'}\Gamma(1+K_{A1})\Gamma(1+K_{B3}) K_{23}^{K_{A1}-K_{B3}}K_{13}^{-1-K_{A1}}U(1+K_{A1},1+K_{A1}-K_{B3}, \kappa-i\epsilon)\\
A_8&=\frac{g_\text{o}^3}{\alpha'}\Gamma(1+K_{A1})\Gamma(1+K_{B3})K_{B2}^{K_{A1}-K_{B3}}K_{B1}^{-1-K_{A1}}U(1+K_{A1},1+K_{A1}-K_{B3}, \kappa-i\epsilon).
\end{align}
However, one must be more careful with Diagram 9, which takes the form 
\begin{align}
&K_{AB}^{-1-K_{A1}}K_{B2}^{K_{A1}-K_{B3}} \Gamma(1+K_{A1})\Gamma(K_{B3}-K_{A1}){_1F_1}(1+K_{A1},1+K_{A1}-K_{B3},\kappa)\notag\\
&\hspace{5 mm}+K_{AB}^{-1-K_{B3}}K_{A2}^{K_{B3}-K_{A1}} \Gamma(1+K_{B3})\Gamma(K_{A1}-K_{B3}){_1F_1}(1+K_{B3},1+K_{B3}-K_{A1},\kappa).
\end{align}
Grouping terms as in (\ref{rearrange}) yields 
\begin{align}
&\Gamma(1+K_{A1})\Gamma(1+K_{B3})K_{AB}^{-1-K_{A1}}K_{B2}^{K_{A1}-K_{B3}} \left(\frac{\Gamma(K_{B3}-K_{A1})}{\Gamma(1+K_{B3})}{_1F_1}(1+K_{A1},1+K_{A1}-K_{B3},\kappa)\notag\right.\\
&\left.\hspace{31 mm}+(\kappa+i\epsilon)^{K_{B3}-K_{A1}}\frac{\Gamma(K_{A1}-K_{B3})}{\Gamma(1+K_{A1})}{_1F_1}(1+K_{B3},1+K_{B3}-K_{A1},\kappa)\right).
\end{align}
It is crucial for obtaining the correct phase that $\kappa$ is taken on the opposite side of the branch cut from the other diagrams. This expression may now be expressed in terms of the $U$ function,
\begin{align}
A_9&=\frac{g_\text{o}^3}{\alpha'}\Gamma(1+K_{A1})\Gamma(1+K_{B3})K_{B2}^{K_{A1}-K_{B3}}K_{AB}^{-1-K_{A1}}U(1+K_{A1},1+K_{A1}-K_{B3}, \kappa+i\epsilon).
\end{align}
\indent Finally, let us dispense with the remaining eight diagrams. It was shown in \cite{zak} that these contain factors either of the form $\exp(-i\pi K_{AB})$ or $\exp(-i\pi K_{23})$. This is analogous to the situation for the ordering $A_{su}$ at four points, which contains an overall factor of $e^{i\pi \alpha' s}$. We see that after introducing shifts $K_{AB}-2 i\alpha'E\Gamma$ and $K_{23}-i\alpha' (E_2+E_3)\Gamma$, such diagrams are subdominant at large $\Gamma$.  We are interested in weak string coupling, for which $\Gamma$ is not large, but still the distinct $\Gamma$ dependence shows that these diagrams do not interfere with the four we are analyzing.  
Relatedly, they contain a large time delay for one of the strings, and cannot interfere with the four diagrams that we have computed.

\begin{figure}
\begin{center}
\begin{tikzpicture}[scale=2]
\draw (-1,-1) -- (-.5,0) -- (-1,1);
\draw (1,-1) -- (.5,0) -- (1,1);
\draw (.5,0) -- (.5,1);
\draw (-.92,.55) node {$k_A$};
\draw (.92,.55) node {$k_3$};
\draw (.38,.75) node {$k_2$};
\draw (.92,-.55) node {$k_B$};
\draw (-.92,-.55) node {$k_1$};
 \path [-,draw,snake it] (-.5,0) -- (.5,0);
\end{tikzpicture}\hspace{30 mm}
\begin{tikzpicture}[scale=2]
\draw (-1,-1) -- (-.5,0) -- (-1,1);
\draw (1,-1) -- (.5,0) -- (1,1);
\draw (-.5,0) -- (-.5,1);
 \path [-,draw,snake it] (-.5,0) -- (.5,0);
 
\draw (-.92,.55) node {$k_A$};
\draw (.92,.55) node {$k_3$};
\draw (-.38,.75) node {$k_2$};
\draw (.92,-.55) node {$k_B$};
\draw (-.92,-.55) node {$k_1$};
\end{tikzpicture}
\end{center}
\caption{\label{separation7}The two contributions to Diagram 7, corresponding to the two terms in (\ref{vertices7}), as in Figure 2 of \cite{BP}.}

\end{figure}
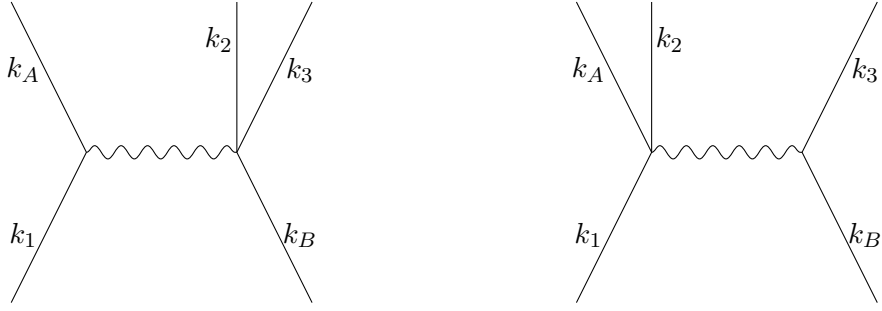

\subsection{Emission location of String 2 and Reggeon analysis}\label{emissionlocation}

It is interesting to ask if there is an interpretation of the division of the amplitude into the two terms in (\ref{approx}), even though a priori it is inconsistent to consider either of the terms in isolation due to the spurious poles at integer values of $K_{B3}-K_{A1}$.  To answer this question, following \cite{BP}\ we rewrite the amplitude in a more suggestive form,
\begin{align}
\Gamma(1+K_{A1})K_{A3}^{-1-K_{A1}}V_3(K_{A1},K_{B3},\kappa)+\Gamma(1+K_{B3})K_{A3}^{-1-K_{B3}}V_A(K_{A1},K_{B3},\kappa)\label{vertices7},
\end{align}
where the vertex functions $V_A$ and $V_3$ are defined as 
\begin{align}
V_3
&=K_{23}^{K_{A1}-K_{B3}}\Gamma(K_{B3}-K_{A1}){_1F_1}(1+K_{A1},1+K_{A1}-K_{B3},\kappa)\nonumber\\
&=\frac{\pi e^{i\pi(K_{B3}-K_{A1})} |K_{23}|^{K_{A1}-K_{B3}}}{\sin(\pi(K_{B3}-K_{A1}))}{_1{\tilde F}_1}(1+K_{A1},1+K_{A1}-K_{B3},\kappa)\\
V_A
&=K_{A2}^{K_{B3}-K_{A1}}\Gamma(K_{A1}-K_{B3}){_1F_1}(1+K_{B3},1+K_{B3}-K_{A1},\kappa)\nonumber\\
&= \frac{\pi K_{A2}^{K_{B3}-K_{A1}}}{\sin(\pi(K_{A1}-K_{B3}))}{_1{\tilde F}_1}(1+K_{B3},1+K_{B3}-K_{A1},\kappa).
\end{align}
Here $_1{\tilde F}_1$ is the regularized confluent hypergeometric function, containing zeros but no poles, and we have used the relation $K_{23}=e^{-i\pi}|K_{23}|$ discussed above.  
The first term in (\ref{vertices7}) contains a single-Regge propagator with momentum transfer $K_{A1}$, times a vertex factor. The single-Regge propagator is the same as in the four-point function, so it is reasonable to conjecture that this term is the amplitude for String 2 to be emitted from String 3, as in Figure \ref{separation7}. Similarly, the second term in (\ref{vertices7}) is the contribution to the amplitude from the case where String 2 is emitted from String A.

The imaginary part of the amplitude comes from the first term here; below we will see that it can be interpreted in terms of an intermediate $s_{23}$-channel string which decays into String 2 and String 3.  
As such, for the imaginary part of the amplitude, the unphysical poles must cancel just within this term.  Indeed,  this can be seen in the second form given above:  the imaginary part of the phase $e^{i\pi(K_{B3}-K_{A1})}$, $~\sin\pi(K_{B3}-K_{A1})$, cancels the unphysical poles in the $\Gamma$ functions.  Thus this term makes sense by itself.  As we will see in the next two sections, the remaining factor   $_1{\tilde F}_1$ oscillates in a way that will be important for the determination of the peak trajectories (impact parameter and time shifts) in the amplitude.      
        
\indent To show more generally that the above interpretation is valid, let us return to the integral expression for the double-Regge amplitude, the left hand side of (\ref{approx}),
\begin{align}
\int_0^\infty dy_2\int_0^\infty dx\,  x^{K_{A1}} y_2^{K_{B3}} \exp(-K_{23}y_2-K_{A2}x-K_{A3}xy_2),
\end{align}
where $x=y_A/y_2$. As explained in \cite{strassler5pt}, the integral contains two dominant contributions, one with $x\ll y_2$ and the other with $y_2\ll x$. In order to isolate the part of the integral where String 2 is emitted from String A, we want to look at the contribution from $y_2\ll x$, corresponding to a short propagator between strings A and 2. Following \cite{strassler5pt}, we change variables from $y_2$ to $v=xy_2$. After rescaling the integration variables, we then find
\begin{align}
K_{A3}^{-1-K_{B3}}K_{A2}^{K_{B3}-K_{A1}}\int_0^\infty dv\int_0^\infty dx\,  x^{K_{A1}-K_{B3}-1} v^{K_{B3}} \exp\left(-x-v-\frac{\kappa v}{x}\right).
\end{align}
We may evaluate this integral as a power series in $\kappa v/x$. This cuts out the small $x$ region of the integral, but keeps the small $y_2$ region, which was our goal. One finds that this power series exactly reproduces the second term in (\ref{vertices7}), as expected.\\
\indent To see that the first term in (\ref{vertices7}) corresponds to radiation from String 3, it is simplest to choose an alternate integral expression for the amplitude,
\begin{align}
\int_0^1 dy_2\int_{0}^{1} dx\, x^{K_{B3}}y_2^{K_{A1}}(1-xy_2)^{K_{A3}}(1-y_2)^{K_{A2}}(1-x)^{K_{23}},
\end{align}
where $x=y_3/y_2$. A short propagator between B and $3$ corresponds to $y_3\ll x$. Repeating the computation of the previous paragraph, we obtain 
\begin{align}
K_{A3}^{-1-K_{A1}}K_{23}^{K_{A1}-K_{B3}}\int_0^\infty dv\, \int_0^\infty dx\, x^{K_{B3}-K_{A1}-1}v^{K_{A1}}\exp\left(-x-v-\frac{\kappa v}{x}\right).
\end{align}
As above, expanding in a power series in $\kappa v/x$ cuts out the small $x$ region of the integral, and we reproduce the first term in (\ref{vertices7}).

\section{Time delays and advances at five points}\label{sec:wavepackets}
In the previous section we computed the five-point amplitude in momentum space, but its interpretation in position space is not yet clear. As in Section \ref{fourpoints}, we are interested in the behavior of the amplitude as a function of the initial impact parameter, as well as the time shifts of the final states. To extract this data from the momentum space amplitude, we will need to introduce a straightforward generalization of the wavepacket analysis in Section \ref{wavepackets}.  We will then combine this with the information from the Regge limit analysis in Section \ref{naive}. \\
\indent We work in the double-Regge regime described in Section \ref{naive}, with String 1 coming out at a small angle $\theta_1$ relative to the trajectory of String A, and similarly for String 3 and String B.  String 2 is the outgoing radiation, whose energy is small compared to the energies of the other strings.  
We will use wavepackets for the outgoing strings 1, 2, and 3 in various combinations to determine their trajectories at future asymptotic infinity. This will enable us to check in particular for {\it advanced} emission, where an outgoing string unambiguously emerges before the putative center of mass collision.
\subsection{Setting up the wavepackets}\label{wavepackets5}
As in Section \ref{wavepackets}, the first step is to choose the initial and final states. For the initial state $|i\rangle$ we will work with the same wavepacket as at four points, namely (\ref{initial}). Suppose that we are interested in the times $T_1$ and $T_2$ at which strings 1 and 2 emerge. For this purpose, it is sufficient to choose the final state
\begin{align}
\langle f|=\int d\tilde{k}_{1x_1}\,d\tilde{k}_{2x_2} \, e^{i\tilde{k}_{1x_1}T_1+i\tilde{k}_{2x_2}T_2}\exp\left(-\frac{(\tilde{k}_{1x_1}-k_{1x_1})^2+(\tilde{k}_{2x_2}-k_{2x_2})^2}{2\sigma_\text{L}^2}\right)\langle \tilde{k}_1,\tilde{k}_2,k_3|,
\end{align}
where $x_1=x\cos\theta_1+y\sin\theta_1$ and $x_2=x\cos\theta_2+y\sin\theta_2$ are the directions of motion of strings 1 and 2. The scattering amplitude of interest is the overlap $\langle f|i\rangle$.\\
\indent Next we need to solve the constraints arising from momentum conservation. Proceeding exactly as in Section \ref{wavepackets}, we have
\bea\label{justsayitmore}
\tilde k_A &=& k_A + (\delta\tilde k_{Ax}, \delta\tilde k_{Ax}, \tilde k_{Ay})+O(\tilde\delta^2)\\
\tilde k_B &=& k_B + (-\delta\tilde k_{Bx}, \delta\tilde k_{Bx}, \tilde k_{By})+O(\tilde\delta^2)\\
\tilde k_1 &=& k_1 + (\delta\tilde k_{1x_1} , \delta\tilde k_{1x_1}\cos\theta_1,\delta\tilde k_{1x_1}\sin\theta_1)+O(\tilde\delta^2).\\
\tilde k_2 &=& k_2 + (\delta\tilde k_{2x_2} , \delta\tilde k_{2x_2}\cos\theta_2,\delta\tilde k_{1x_1}\sin\theta_2)+O(\tilde\delta^2).
\eea
where the central values $k_I^\mu$ of the momenta are given in Appendix \ref{kinematics}.
Energy-momentum conservation gives
\begin{align}
\delta \tilde{k}_{Ax}&=-\frac{1}{2}(1+\cos\theta_1)\delta \tilde{k}_{1x_1}-\frac{1}{2}(1+\cos\theta_2)\delta \tilde{k}_{2x_2}+O(\tilde{\delta}^2)\\
\delta \tilde{k}_{Bx}&=\frac{1}{2}(1-\cos\theta_1)\delta \tilde{k}_{1x_1}+\frac{1}{2}(1-\cos\theta_2)\delta \tilde{k}_{2x_2}+O(\tilde{\delta}^2)\\
\tilde{k}_{+}&=-\delta \tilde{k}_{1x_1}\sin\theta_1-\delta \tilde{k}_{2x_2}\sin\theta_2+O(\tilde{\delta}^2).
\end{align}
The amplitude then becomes
\begin{align}\int d\delta \tilde{k}_{1x_1}\,d\delta \tilde{k}_{2x_2}\,d \tilde{k}_{-}\, e^{-i\tilde{k}_{-}b/2+i\delta \tilde{k}_{1x_1}T_1+i\delta \tilde{k}_{2x_2}T_2}\exp\left(-\frac{\delta \tilde{k}_{1x_1}^2}{2\sigma_{\text{L},\text{eff},1}^2}-\frac{\delta \tilde{k}_{2x_2}^2}{2\sigma_{\text{L},\text{eff},2}^2}-\frac{\tilde{k}_{-}^2}{2\sigma_\text{T}^2}\right)A(\tilde{K}_{IJ}).
\end{align}
The effective widths $\sigma_{\text{L},\text{eff},1}$ and $\sigma_{\text{L},\text{eff},2}$ are functions of the longitudinal and transverse widths as in Section \ref{wavepackets}. \\
\indent In order to compute the peak impact parameter and time shifts, we need to isolate the rapidly oscillating part of the amplitude. In the four diagrams that we will consider, the amplitude takes the form 
\begin{align}
A(K_{IJ})=\exp(i\delta(K_{A1},K_{B3}))A_{\text{slow}}(K_{IJ}).
\end{align}
The phase is only a function of the fixed variables $K_{A1}$ and $K_{B3}$. The leading dependence of these kinematic invariants on the integration variables is the following, including the deformations $\tilde k=k+\delta\tilde k$ away from the central values of the momenta:
\begin{align}
\alpha'^{-1}\tilde{K}_{A1}&\sim 2EE_1(1-\cos\theta_1)-2E(1-\cos\theta_1)\delta \tilde{k}_{1x_1}-E_1\sin\theta_1 \tilde{k}_{-}\notag\\
&\hspace{5 mm}-E_1(1-\cos\theta_1+\cos\theta_2-\cos(\theta_1-\theta_2))\delta \tilde{k}_{2x_2}\\
\alpha'^{-1}\tilde{K}_{B3}&\sim 2EE_3(1-\cos\theta_3)-E_3(1-\cos\theta_1+\cos(\theta_1-\theta_3)-\cos\theta_3)\delta \tilde{k}_{1x_1}-E_3\sin\theta_3 \tilde{k}_-\notag\\
&\hspace{5 mm}-E_3(1-\cos\theta_3+\cos(\theta_2-\theta_3)-\cos\theta_2)\delta \tilde{k}_{2x_2}\label{tildek}.
\end{align}
The phase is then stationary when 
\begin{align}
b=2\frac{\partial\delta}{\partial\tilde{k}_-}\hspace{15 mm}T_1=-\frac{\partial \delta}{\partial \delta \tilde{k}_{1x_1}}\hspace{15 mm}T_2&=-\frac{\partial \delta}{\partial \delta \tilde{k}_{2x_2}}.
\end{align}
It is useful to express this condition in terms of derivatives with respect to the kinematic invariants $K_{IJ}$. From (\ref{tildek}) we find that
\begin{align}\label{peaks5pts}
\alpha'^{-1}b&=-2E_1\sin\theta_1\frac{\partial\delta}{\partial{K}_{A1}}-2E_3\sin\theta_3\frac{\partial\delta}{\partial {K}_{B3}}\\
\hspace{-3 mm}\alpha'^{-1}T_1&=2E(1-\cos\theta_1)\frac{\partial \delta }{\partial {K}_{A1}}+E_3(1-\cos\theta_1+\cos(\theta_1-\theta_3)-\cos\theta_3)\frac{\partial \delta}{\partial {K}_{B3}}\\
\alpha'^{-1}T_2&=E_1(1-\cos\theta_1+\cos\theta_2-\cos(\theta_1-\theta_2))\frac{\partial \delta }{\partial {K}_{A1}}\notag\\
&\hspace{5 mm}+E_3(1-\cos\theta_3+\cos(\theta_2-\theta_3)-\cos\theta_2)\frac{\partial \delta }{\partial {K}_{B3}}\label{peaks5ptsend}.
\end{align}

With more general wavepackets, we can similarly localize the strings in both $x$ and $y$ (equivalently in the rotated directions $x_1=x\cos\theta_1+y\sin\theta_1$ and $y_1=-x\sin\theta_1+y\cos\theta_1$), completely determining their peak trajectories in the 2+1 dimensions of our kinematics.  We will implement this below after determining the phases which contribute from our calculated amplitude.  

\subsection{Phases of the amplitudes}
The analysis of the previous section yields the time shifts and peak impact parameters for a given phase of the amplitude. The only remaining step is to extract these phases from the four diagrams under consideration at five points. Let us start with Diagram 7, which takes the form (\ref{Seven}). The phase of the dependence of the two Regge propagators is
\begin{align}
(e^{-i\pi}|K_{23}|)^{K_{A1}-K_{B3}}K_{A3}^{-1-K_{A1}}=e^{i\pi (K_{B3}-K_{A1})}|K_{23}|^{K_{A1}-K_{B3}}K_{A3}^{-1-K_{A1}}.\label{propphase}
\end{align}
Note that this is the same phase as the first term in (\ref{approx}). If one could neglect the vertex function $U(a,b,z)$, then this would be the full phase of the amplitude. \\
\indent In fact, the function $U(a,b,z)$ does have a nontrivial phase for generic values of its parameters. To see this, let us apply the transformation 
\begin{align}
U(a,b,z)=z^{1-b}U(1+a-b,2-b,z)
\end{align}
to the amplitude $(\ref{Seven})$. We find
\begin{align}\label{sevenotherform}
A_7&= \frac{g_\text{o}^3}{\alpha'}\Gamma(1+K_{A1})\Gamma(1+K_{B3})K_{A2}^{K_{B3}-K_{A1}}K_{A3}^{-1-K_{B3}}U(1+K_{B3},1+K_{B3}-K_{A1}, \kappa-i\epsilon).
\end{align}
Since the kinematic invariants $K_{A2}$ and $K_{A3}$ are positive, the phase of the two Regge propagators is now trivial, as in the second term in (\ref{approx}). It follows that $U$ must generically contribute a nontrivial phase.\\
\indent This means that to obtain definite results for the data $b$, $T_1$, and $T_2$, we must  restrict the analysis to a range of parameters. Let us choose to take $K_{B3}>K_{A1}$, while keeping both $K_{A1}$ and $K_{B3}$ much smaller than all $s$-like variables. Since the momentum transfer $K_{B3}$ is larger than $K_{A1}$, it is natural to conjecture that in this limit String 2 is emitted from String B or String 3, meaning that the first term in the decomposition (\ref{vertices7}) determines the phase.  As discussed below that equation, the imaginary part only gets a contribution from this first term. In Appendix \ref{numerics} we provide numerical evidence that this is indeed the case when $\sin\theta_1$ and $\sin\theta_2$ have the same sign, so that the phase is as given in (\ref{propphase}).  The phases of each of the diagrams are then 
\begin{align}\label{phasesB3}
e^{i\delta_1}&=e^{i\pi K_{B3}}\\
e^{i\delta_7}&=e^{i\pi(K_{B3}-K_{A1})}\\
e^{i\delta_8}&=1\\
e^{i\delta_9}&=e^{i\pi K_{A1}}\label{phasesB3end}.
\end{align}
In the opposite regime $K_{A1}>K_{B3}$ we instead find that when $\sin\theta_3$ and $\sin\theta_2$ have the same sign, the phases are
\begin{align}
e^{i\delta_1}&=e^{i\pi K_{A1}}\label{phasesA1start}\\
e^{i\delta_7}&=1\\
e^{i\delta_8}&=e^{i\pi(K_{A1}-K_{B3})}\\
e^{i\delta_9}&=e^{i\pi K_{B3}}\label{phasesA1}.
\end{align}
\indent These phases satisfy various consistency conditions. For example, note that switching $k_A\leftrightarrow k_1$ and $k_B\leftrightarrow k_3$ is equivalent to switching Diagram 1 with Diagram 9, and Diagram 7 with Diagram 8. The phases should therefore remain invariant if we apply both of these permutations, and we see from (\ref{phasesB3})-(\ref{phasesA1}) that this is indeed the case.\\
\indent Another important check is that each of the phases should reduce to those of the corresponding ordering at four points. If we remove String 2, then the topologies of Diagrams 1 and 9 both reduce to the ordering $A_{st}$, whose phase is $e^{-i\pi \alpha' t}$, while the topologies of Diagrams 7 and 8 reduce to the ordering $A_{tu}$, which has trivial phase. The results (\ref{phasesB3})-(\ref{phasesA1}) indeed have this limiting behavior as $E_2\to 0$, or equivalently $K_{A1}\to K_{B3}$.
\subsection{Diagram 7: Early interaction}
Now that we have computed the phases of all of the diagrams, we are ready to analyze the peak time shifts and impact parameters, starting with Diagram 7. This diagram is the sum of two terms, with phases 1 and $\exp(i\pi (K_{B3}-K_{A1}))$. As long as $K_{B3}>K_{A1}$ and $\sin\theta_1,\sin\theta_2>0$, the results of Appendix \ref{numerics} show that the phase is $\exp(i\pi(K_{B3}-K_{A1}))$. This phase corresponds to the production of long strings which decay into strings 2 and 3. The results for the impact parameter and time shifts are
\begin{align}
b&=-2\pi \alpha' E_2\sin\theta_2\\
T_1&=\pi \alpha'E_2\theta_1\sin\theta_2-\frac{\pi\alpha'}{2}(1+\cos\theta_2)\theta_1^2E_2\\
T_2&=2\pi\alpha' E\theta_1\sin\theta_2+\pi\alpha' E_2\sin^2 \theta_2.
\end{align}
\indent Let us first examine the result for $T_1$, which says that the asymptotic trajectory of String 1 satisfies $x_1=T-T_1$. It is convenient to consider this for a moment in a shifted coordinate system with String A aimed to hit $x_1=0$ at $T=0$, which is to say $y_A=0$.   The time shift for String 1 in the shifted coordinates is
\begin{align}\label{advance}
T_1+\frac{b}{2}\sin\theta_1=-\frac{\pi \alpha'}{2}(1+\cos\theta_2)\theta_1^2E_2,
\end{align}
which is negative.  This will lead us to a time advance for String 1, meaning that it emerges from the scattering process earlier than it would have if the strings had begun to scatter at the location of the center of mass collision.
To see this we will put together the data obtained from the scattering process; we will stick to the original (unshifted) coordinate system for the remainder of our discussion.      
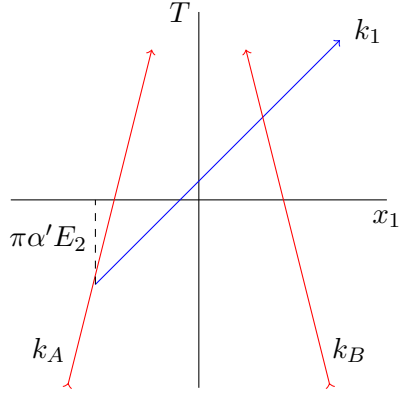
\begin{figure}
\begin{center}
 \begin{tikzpicture}[scale=2.5]
\draw(0,-1) -- (0,1);
\draw (-1,0) -- (1,0);
\draw[->,color=blue] (-.55,-.45) -- (.75,.85); 
\draw (.9,.9) node{$k_1$};
\draw[>->,color=red] (.7,-1) -- (.25,.8);
\draw[>->,color=red] (-.7,-1) -- (-.25,.8);
\draw (1,-.1) node{$x_1$}; 
\draw (-.1,1) node{$T$}; 
\draw (-.8,-.8) node{$k_A$}; 
\draw (.8,-.8) node{$k_B$}; 
\draw[dashed] (-.55,-.45) -- (-.55,0);
\draw (-.8,-.2) node{$\pi \alpha' E_2$};
\end{tikzpicture}
\end{center}
\caption{ \label{1advanceunambiguous} The dynamics of Diagram 7 projected on the plane of $T$ and $x_1=x\cos\theta_1+y\sin\theta_1$. 
As described in the text, the dominant contribution to the amplitude derived in Section \ref{naive}\ indicates that 1 emerges directly from A, after $A$ emits a Reggeon.   The wavepacket analysis of trajectories shows that this is consistent:  the traced-back trajectory of 1 directly intersects that of A. This, combined with the absence of a transverse-spreading wavefunction factor, indicates that the process is as depicted, requiring an early interaction.  
}
\end{figure}

\indent  It is worthwhile to record additional data from the amplitude on the trajectories; in particular we can obtain more information about the outgoing trajectory of String 1 by simply applying additional wavepackets to localize its transverse coordinate $y_1=-x\sin\theta_1+y\cos\theta_1$ at some position $y_1=b_1$. First let us see what value of $b_1$ we expect to find, if the trajectory of String 1 is the continuation of the trajectory of String A. The asymptotic path of String A satisfies $x=T, y=-\pi E_2\alpha'$, equivalently 
\begin{align}
(x_1,y_1)=(T\cos\theta_1-\pi \alpha' E_2\sin\theta_2\sin\theta_1,-T\sin\theta_1-\pi \alpha' E_2\sin\theta_2\cos\theta_1).
\end{align}
Since String 1 flies out along the $x_1$ direction, it follows a path of the form 
\begin{align}
(x_1,y_1)=(T-T_1,b_1).
\end{align}
for a constant value of $b_1$ which we will determine momentarily.  
First, setting these equal and solving for $b_1$ gives
\begin{align}\label{peakimpact1}
b_1&=-\pi \alpha' E_2\sin \theta_2+\pi \alpha' E_2\theta_1(1+\cos\theta_2).
\end{align}
To check that this is indeed the peak value of $b_1$, we compute the scattering amplitude with a state for String 1 of the form
\begin{align}
\langle f|=\int d\tilde{k}_{1x_1}\, d\tilde{k}_{1y_1} e^{i\tilde{k}_{1x_1}T_1-i\tilde{k}_{1y_1}b_1}\exp\left(-\frac{(\tilde{k}_{1x_1}-k_{1x_1})^2}{2\sigma_\text{L}^2}-\frac{(\tilde{k}_{1y_1}-k_{1y_1})^2}{2\sigma_\text{T}^2}\right)\langle \tilde{k}_1|\label{transverse5pts},
\end{align}
Solving momentum conservation and using the saddlepoint approximation as in Sections \ref{wavepackets} and \ref{wavepackets5}, one finds that (\ref{peakimpact1}) is satisfied.  In terms of the original coordinates, the trajectories of String A and String 1 intersect at $(T, x, y)=(-\pi\alpha' E_2, -\pi\alpha' E_2, -\pi\alpha' E_2)$.  
Using wavepackets to localize strings B and 3, we similarly find that the traced-back trajectory of String 3 exhibits a delay, consistent with its emerging from String B at $(T, x, y)=(\pi\alpha' E_2, -\pi\alpha' E_2, \pi\alpha' E_2)$.

According to the analysis in Appendix \ref{naive}, the trajectory of String 1 is the continuation of the trajectory of String A in small-angle Regge scattering.  As we have just seen from our wavepacket analysis folded against the amplitude that we (re-)derived above, the traced-back peak trajectories of strings A and 1 intersect directly, as do strings B and 3, showing that this interpretation is consistent.  

The final element we need is an estimate of the effects of the transverse spreading of the string, at the level that can be detected in a scattering amplitude of center of mass energy $E$.\footnote{We thanks S. Giddings for extensive discussions.}  Although the peak trajectories
of String A and String 1 are separated by $\Delta y=\pi\alpha' E_2\theta_1$ at time $T=0$, could it be that the 
interaction proceeds via a transverse effect which occurs at $T=0$, rather than via the turning of String A into String 1 early as indicated by the central trajectories for the strings derived above?  
This would require not only that there exists transverse spreading of the string, but that a transversely-separated emission of 1 from A {\it dominates} over a process where A simply turns into 1 (at the point where it emits the Reggeon).

From the discussion in Section \ref{longvstrans}, we see that any interaction which arises as a result of transverse spreading by a distance $\Delta y$ should come with an additional penalty factor in the amplitude of the form $\exp(-(\Delta y)^2/(\alpha' \log( E/E_0)))$. In particular, if 1 emerged not from the center $y_A=-\pi E_2\alpha'$ 
of A's trajectory, but instead from the tail of A's transverse spreading wavefunction,
 out at some distance $y_A+\Delta y$, 
we would expect an extra suppression factor in our amplitude convolved with wavepackets that
localize the trajectories: a factor
 of the form $\exp(-(\Delta y)^2/(\alpha' \log( E/E_0)))$.  
The amplitude does not exhibit a corresponding suppression factor.  This was also the case for the four-point amplitude with a nonzero peak impact parameter.             

So finally, given that 1 emerges from the center of A, it follows from causality that there must have been an early interaction, as shown in Figure \ref{1advanceunambiguous}. This effect comes from the stringy, Regge factor in the amplitude and would not arise in a theory of point particles. Later in this section we will give a lower bound on the extent of the longitudinal nonlocality.\\
\indent It is straightforward to determine the outgoing trajectory of String 2 as well. Let us take $\theta_2=\pi/2$ for simplicity, and localize String 2 at $x=b_2$. The amplitude is peaked at 
\begin{align}
T_2&=\pi \alpha' (E_2+2E\theta_1)\\
b_2&=-\frac{\pi \alpha'}{2}\left(\frac{E_2^2}{E}+2E_2\theta_1+2E\theta_1^2\right)\approx 0.
\end{align}
This trajectory does not intersect any of the asymptotic trajectories of the other strings, which is consistent with the picture that String 2 is emitted from the decay of an intermediate long string instead of directly from one of the incoming or outgoing states.

\subsection{Diagram 9: Late Bremsstrahlung from the rhombus}

As mentioned above in Figure \ref{5ptpics}, another use of the radiation leg (String 2) is to make an additional check of the direct bending of String A into String 1 and String B into String 3 by looking for radiation emitted from the turning points. 
Diagram 9 has an ordering which reduces to that of the $A_{st}$ diagram at four points.
The phase in the regime $K_{B3}>K_{A1}$ and $\sin\theta_1,\sin\theta_2>0$ is $e^{i\pi K_{A1}}$, which agrees with the phase of the contribution to the worldsheet integral from $y_2\sim y_B$. Therefore we expect to see Bremsstrahlung emitted from the long string at the turning point of $B$, as shown in Figure \ref{5ptpics}.\\
\indent Repeating the analysis of the previous section, one finds for $E_2\ll E\theta_1$, the trajectory of String 2 intersects the $y$-position of String B at 
\begin{align}
(T,x,y)=\left(\frac{\pi \alpha' E\theta_1^2}{2},-\frac{\pi \alpha' E\theta_1^2}{2},\pi \alpha'E\theta_1\right).
\end{align}
As anticipated, this is exactly the point where String B turns in Figure \ref{5ptpics}, so the trajectory of String 2 is consistent with radiation from the corner of the rhombus. It is also straightforward to check that the trajectory of 1 is the continuation of the trajectory of A, which further corroborates the early interaction that we have derived for Diagram 7. \\
\indent In this diagram  (Diagram 9), the only on-shell single-string state that can be created is between $k_1$ and $k_3$.  (One can also run the process time-reversed, obtaining absorption of 2 at the turning point of 3 into B.) This makes it straightforward to treat String 2 as a perturbation, since it does not introduce additional on-shell poles.
For $E_2\ll E\theta_1$, this on-shell state should be the same long string that is made at four points in the ordering $A_{st}$.  
Diagram 1 also reduces to $A_{st}$ at four points, but at five points it is slightly more involved, since it is kinematically possible to make long strings between A and B, between $1$ and $2$, or between 2 and 3. It would be interesting to understand if there is a classical picture of these strings.  In general, it would also be interesting to analyze further the other diagrams, although the only diagram with an obvious early effect is Diagram 7.\footnote{However, we have not yet performed an exhaustive analysis in all single and double Regge regimes, so additional effects of interest may be present in the S-matrix data.} 

\subsection{Extent of the longitudinal nonlocality and string spreading}

\indent The time advance (\ref{advance}) is small in the Regge limit $E\gg 1/\sqrt{\alpha'}$, $\theta_1\ll 1$, $E_2\ll E$. However, from the geometry we have derived above, it indicates a level of advanced longitudinal spreading that is at least of order $\alpha' E_2$. 
\begin{align}
(T,x,y)=\left(-\pi \alpha' E_2,-\pi\alpha'E_2,-\pi \alpha' E_2\right).
\end{align}
where its trajectory matches the traced-back trajectory of String 1.  
If String A then turns immediately to move in the $x_1$ direction, this produces the time advance that we have computed for String 1.  The trajectory joining A and 1 may be smoother than this, of course, but if it began deviating from a straight line later than $T=-\pi \alpha' E_2$ then there would be no way for it to causally reproduce the outgoing trajectory. Given our assumptions, this implies that String A begins to interact at least as early as a time $-\pi\alpha'  E_2$. 

\indent Moreover, we can remain in the regime where the Regge approximation is valid, and still obtain a much larger time advance, albeit at substantial cost in the magnitude of the scattering amplitude as we ramp up the energy $E_2$ and the scattering angle $\theta_1$.  Parametrically, the time advance could be as large as of order $\epsilon\alpha' E$, with $\epsilon$ a control factor that is sufficiently smaller than 1 so that the Regge analysis of our amplitude is valid.   The limiting possibility, $T_{1}\to -\alpha' E$, is the extent of longitudinal spreading predicted by light cone calculations  \cite{lennyspreading}.  
It is worth noting that the interpretation of the four-point amplitude in terms of longitudinal spreading in Section \ref{longvstrans} requires a longitudinal spreading scale of $\sim \alpha' E$ as predicted in \cite{lennyspreading}, without a similar degradation of the amplitude.  

If such spreading is detected with sufficient amplitude,  it produces sufficient `drama' for a late-infaller in black hole physics \cite{usBH} to potentially address recent puzzles \cite{firewalls}\ explicitly via string theoretic corrections to the naive effective field theory description of the dynamics of a late probe of a black hole \cite{backdraft}.

\section{Conclusions}

In this work we have found that the peak trajectories, intermediate string solutions, Bremsstrahlung radiation, and Reggeon physics all point consistently to a rather simple spacetime geometry for tree-level string amplitudes.  The geometry requires some degree of longitudinal nonlocality, if we assume the more standard transverse string spreading as reviewed and refined here and in \cite{usBH}.  

There are many directions for future work.
First, although we have gathered extensive S-matrix `data' so far in our study, we have not performed an exhaustive analysis in all regimes.    
It be interesting to study other regimes at five points, such as single-Regge limits \cite{BP}, which are still tractable and less suppressed in amplitude than the double-Regge limits on which we have focused in the present work.  

More generally, it will be interesting to investigate higher point amplitudes, setting up an experiment where two relatively boosted strings are created spatially separated, to avoid a direct center of mass collision.  Their interaction would require nonlocality, in a setup that is closer to the situation for black hole infallers in \cite{usBH}.\footnote{We thank Don Marolf for discussions of this possibility.}  
And as mentioned in \cite{usBH}, generalizations to scattering in AdS/CFT may provide additional insight, intermediate between the present flat space S-matrix elements and full-fledged black hole geometries.

\section*{Acknowledgements}
We thank S. Giddings for many insights during our intermediate collaboration.  We are also grateful to  
Thomas Bachlechner, Liam Fitzpatrick, David Gross,  Daniel Harlow, Juan Maldacena, Don Marolf, Liam McAllister,  Joe Polchinski, Mukund Rangamani, Steve Shenker, Douglas Stanford, Lenny Susskind, and Gabriele Veneziano for very useful discussions over the course of this work. We thank S. Giddings, D. Harlow, J. Maldacena, A. Puhm, and D. Stanford for comments on the manuscript.  This research was supported in part by the National Science Foundation under Grant No. NSF PHY11-25915. The work of E.S.~was supported  in part by the National Science Foundation
under grant PHY-0756174 and NSF PHY11-25915 and by the Department of Energy under
contract DE-AC03-76SF00515. M.D. is supported by a Stanford Graduate Fellowship and a KITP Graduate Fellowship.

\appendix 
\section{Five-point kinematics}\label{kinematics}
In this appendix we will review the kinematics of the five-point interaction in the center of mass frame for the initial strings. The momenta are
\begin{align}\label{fivemomenta}
k_A&=(E,E,0)\\
k_B&=(E,-E,0)\\
k_1&=(-E_1,-E_1\cos\theta_1,-E_1\sin\theta_1)\\
k_2&=(-E_2,-E_2\cos\theta_2,-E_2\sin\theta_2)\\
k_3&=(-E_3,E_3\cos\theta_3,E_3\sin\theta_3).
\end{align}
We can use momentum conservation to solve for the angle $\theta_3$ and energy $E_3$, 
\begin{align}
E_3\sin\theta_3&=E_1\sin\theta_1+E_2\sin\theta_2\\
E_3&=2E-E_1-E_2.
\end{align}
The on-shell condition for $k_3$ determines the energy 
\begin{align}
E_1&=\frac{2E(E-E_2)}{2E+E_2(\cos(\theta_1-\theta_2)-1)}\sim   E -\frac{E_2}{2}(1+\cos(\theta_1-\theta_2)).
\end{align}
where we used $E_2\ll E$ as in (\ref{Etwo}).  For generic $\theta_2$ we can neglect $\theta_1\ll \theta_2$ here, working in the Regge regime for the underlying four-point amplitude with $\theta_1\ll 1$.
 
It is useful to record the behavior of the 10 quantities $K_{IJ}=2\alpha' k_I\cdot k_J$ which enter directly into the string amplitudes.  This is a redundant description, but useful in the calculation.  They come in three sets of variables, the first of which is
\begin{align}\label{KIJs}
  K_{AB} &= -4\alpha' E^2  \\
K_{A3} &\sim 4\alpha' E^2  \\
K_{B1} &= 2\alpha'E E_1(1+\cos\theta_1) \sim 4\alpha' E^2  \\
K_{13} &\sim -4\alpha' E^2.
\end{align}
These four kinematic invariants are of order $\pm s_4$, where $s_4$ is the Mandelstam $s$ variable in the four-point amplitude.  \\
\indent The next set of variables are $t_4$-like,
\begin{align}\label{KIJt}
K_{A1} &= 2\alpha' E E_1(1-\cos\theta_1) \sim 2\alpha' E^2(1-\cos\theta_1)\\
K_{B3} &= \frac{2 \alpha' E ((E_2-2 E) \sin (\theta_1/2)+E_2 \sin
( \theta_1/2-\theta_2))^2}{{E}_2 \cos (\theta_1-\theta_2)-{E}_2+2 E}\sim \alpha'(2E\sin(\theta_1/2)+E_2\sin\theta_2)^2.
\end{align}
These variables are fixed as $E\to\infty$.\\
\indent Finally we have variables which scale like $EE_2$ (hence growing like $E\sim \sqrt{s_4}$ for fixed $E_2$),
\begin{align}\label{KIJtwo}
K_{A2} &= 2\alpha' E E_2(1-\cos\theta_2)  \\
K_{B2} &= 2\alpha' E E_2(1+\cos\theta_2) \\ 
K_{12} &\sim  -2\alpha' E E_2 (1-\cos\theta_2)+\alpha' E_2^2 \sin^2 \theta_2 \\
K_{23} &\sim -2\alpha'E E_2(1+\cos\theta_2)-\alpha'E_2^2\sin^2\theta_2.
\end{align}
\section{Numerical analysis of $U(a,b,z)$}\label{numerics}

In this appendix we will investigate the oscillatory behavior of Diagram 7, using the form (\ref{sevenotherform}).  The phase in this expression comes entirely from the Tricomi confluent hypergeometric function 
\be\label{Ufornum}
U(1+\tilde K_{B3}, 1+\tilde K_{B3}-\tilde K_{A1}, \tilde\kappa -i\epsilon)
\ee
where $\tilde K_{IJ}$ refers to the kinematic invariants perturbed about the central values of the momenta, as in the wavepacket analysis in the main text.  Taking $\theta_2=\pi/2$ for simplicity, we work in the regime $\sin \theta_1>0$ and $K_{B3}>K_{A1}$, and check that the phase is approximately $e^{i\pi(K_{B3}-K_{A1})}$ as described in the text. We will work in coordinates where String A is localized at $y=0$.  

Specifically, we analyze the phase of the amplitude using the following kinematics, where the tilded variables are varied (within a small range dictated by the wave packet width $\sigma$ discussed in the text):
\begin{align}\label{fivemomentatildeone}
\tilde k_A^\mu&=(\sqrt{(E+\delta\tilde{k}_{Ax})^2+\tilde{k}_{Ay}^2},E+\delta\tilde{k}_{Ax},\tilde{k}_{Ay})
\\
\tilde k_B^\mu&=(\sqrt{(E-\delta \tilde{k}_{Bx})^2+\tilde{k}_{By}^2},-E+\delta \tilde{k}_{Bx},\tilde{k}_{By})\\
\tilde k_1^\mu&=(- E_1-\delta \tilde E_1, -(E_1+\delta \tilde E_1)\cos\theta_1, -(E_1+\delta \tilde E_1)\sin\theta_1)\\
\tilde{k}_2^\mu&=k_{2}^\mu\approx (-E_2,0,-E_2)\\
\tilde{k}_{3}^\mu&=k_3^\mu.
\end{align}
Here we work for simplicity in a regime $\theta_2\approx \pi/2$. 

\begin{figure}[htbp]
\begin{center}
\includegraphics[width=8cm]{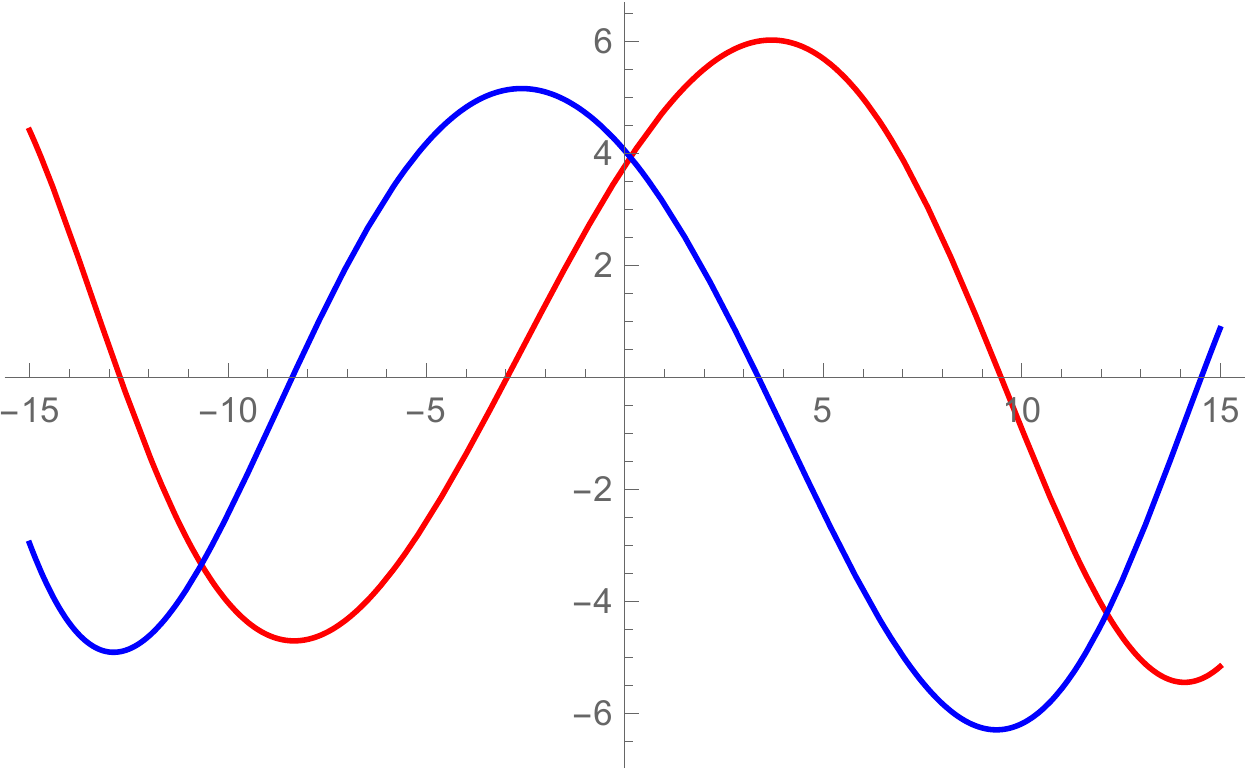} 
\end{center}
\caption{\label{plotrealimag}  
Real (red) and imaginary (blue) parts of $F_\text{{env}}U$ plotted against the variation $\delta\tilde E_1$ of the energy of String 1 integrated over in its wavepacket, in a range between $-E/10$ and $+E/10$.  Parameters are (with $\alpha'=1$) $E=150, ~ E_2=10, ~ \theta_1=1/5\sqrt{3}\approx .1,  ~ \theta_2=\pi/2$.   These exhibit an advance of the scale discussed in the text. } 
\end{figure}

The integrals over $\delta \tilde k_{Ax}, \delta \tilde k_{Bx},$ and $\tilde k_{Ay}$ can be done with the energy-momentum conserving delta function $\delta(\tilde k_A+\tilde k_B+k_1+\tilde k_2+k_3)$.  
Multiplying (\ref{Ufornum}) by a non-oscillating envelope function proportional to 
\be
F_{\text{env}}\propto e^{-\tilde K_{A1}},
\ee    
we obtain in the results for the real and imaginary parts of the function in Figure \ref{plotrealimag}, as a function of $\delta\tilde E_1$.  The relation between the real and imaginary parts as well as the period are correct for a time advance which is approximately $\alpha' E_2\pi\theta_1^2/2$ arising from the phase $e^{i\pi (K_{B3}-K_{A1})}$.  \\
\indent One can similarly plot the dependence of the amplitude on $\tilde{k}_{By}$ (or on $\delta\tilde{k}_{1y_1}$ as in (\ref{transverse5pts})), and the oscillations are again consistent with a phase $e^{i\pi(K_{B3}-K_{A1})}$.

\section{Gaussian deviations from the center of the wavepackets}\label{gaussian}
In the main body of this work we have only considered linear fluctuations of the momenta about the center of the wavepackets, which determine the peak impact parameters and time shifts. Here we will consider this procedure more carefully, and explain the constraints that the widths of the wavepackets must satisfy to justify neglecting the Gaussian fluctuations.\\
\indent Let us repeat the computation of Section \ref{wavepackets}, this time keeping terms up to quadratic order in the integration variables. Since we are mainly interested in the transverse wavepackets here, we will take Particle 1 to be in a momentum eigenstate and set $\delta \tilde{k}_{1x_1}=0$. The solutions to momentum conservation become
\begin{align}
\delta \tilde{k}_{Ax}=-\delta \tilde{k}_{Bx}=-\frac{\tilde{k}_-^2}{8E},\hspace{10 mm}\tilde{k}_+=0.
\end{align}
We can now express the Mandelstam invariant $t$ in terms of the remaining integration variables,\begin{align}
\tilde{t}&=-2E^2(1-\cos\theta)+E\sin\theta\tilde{k}_--\frac{1}{4}\tilde{k}_-^2\cos\theta\sim -\frac{1}{4}(\tilde{k}_--2E\theta)^2.
\end{align}
Here we have expanded for small $\theta$. \\
\indent To illustrate the point that we are trying to make, let us consider the open string ordering $A_{st}$. For $-t\gg 1/\alpha'$, the scattering amplitude becomes
\begin{align}
\int d\tilde{k}_-\, e^{-i(b\tilde{k}_-/2+\pi \alpha'  \tilde{t})}(-\tilde{t})^{-3/2}e^{\alpha' \tilde{t}}\exp\left(-\frac{\alpha' }{4}(\tilde{k}_--2E\theta)^2\log(s/\tilde t)-\frac{\tilde{k}_-^2}{2\sigma_{\text{T}}^2}\right).
\end{align}
Depending on the size of $\sigma_{\text{T}}$, this integral is peaked at different values of $\tilde{k}_-$. For $\sigma_{\text{T}}\ll (\alpha' \log s)^{-1/2}$, the integral is peaked at $\tilde{k}_-=0$, and in the opposite regime the integral is peaked at $\tilde{k}_-=2E\theta$, or $\tilde{t}=0$. The latter is the case of forward scattering, where the momentum $\tilde{k}_A$ aligns with $k_1$. Since we are trying to set up a scattering experiment at finite angle, we certainly do not want the integral to be peaked at forward scattering. The fact that the strings have a logarithmic transverse size ruins our attempt to accurately localize the momenta in the regime $\sigma_{\text{T}}\gg (\alpha' \log s)^{-1/2}$.  Therefore we take $\sigma_{\text{T}}\ll (\alpha' \log s)^{-1/2}$, and the computation goes through as in Section \ref{wavepackets}.\\
\indent At five points the situation works similarly. As long as the transverse momentum space widths are taken sufficiently small, the integral is peaked when the fluctuations around the centers of the wavepackets are small, and one can safely neglect the quadratic terms.

\section{Form factors, wavefunctions, and string amplitudes}\label{transverseappendix}

One approach to determining the extent of the string might be to obtain a distribution of string in the transverse and longitudinal directions as the Fourier transform of a form factor.\footnote{We thank Steve Giddings for extensive discussions of this approach and its connection to \cite{bpst}.}

To begin, let us review the structure of coherent scattering amplitudes, written in terms of form factors \cite{formfactors}.  In standard quantum mechanics in the Born approximation, the scattering amplitude off a source with wavefunction $\Psi_\text{s}(\vec r)$,  as a function of momentum transfer $q=k_1-k_A\approx (0,\vec q)$ takes the form
\be\label{genform}
A(\vec q) =A_{\text{point}}(\vec q)\times F(\vec q).  
\ee
where $A_{\text{point}}$ is the amplitude for scattering off of a point source, and $F$ is a form factor, the Fourier transform of the source density $\rho(\vec r)=|\Psi_{\text{s}}(\vec r)|^2$:
\be\label{Frho}
F(\vec q) = \int d\vec r \, e^{i \vec{q}\cdot \vec{r}} \rho(\vec r).
\ee 
Note that this expression, although approximate, is quantum mechanical and coherent (see \cite{formfactors}\ for an explicit review in the context of atomic and nuclear physics, with the first reference explicitly presenting the coherent cross section in terms of $F$).  Turning this around, if we scatter off a fixed, time-independent source from all directions, and if we also are given the amplitude $A_{\text{point}}$ to scatter off of a point in the source, then we can invert the scattering amplitude to obtain the density $\rho(\vec r)$ of the source.

It is not trivial to generalize this to string scattering, for several reasons.  Before proceeding, note that the scattering we are discussing involves incoming motion along a single direction (the direction of relative motion of our strings A and B).  Even in a simple quantum mechanics problem such as we just reviewed, if we did not scatter from all directions we would not have enough information to determine the distribution in all directions $\vec r$; the directions transverse to the direction of relative motion would be relatively straightforward to determine using forward scattering, but not the distribution in the longitudinal direction.   

Next, in attempting to generalize (\ref{genform}) to string theory, the first factor $A_{\text{point}}$ is more subtle than it is in the above quantum mechanics problem.  This would represent the scattering amplitude of one string off of a point in the other, not something for which we have a well-defined expression a priori.
If we had such an expression for $A_{\text{point}}$, we could pull it off and then Fourier transform the remainder of the forward scattering amplitude to obtain a candidate expression for $\rho(\vec x_{\perp})$ (where $\vec x_\perp$ is transverse to the forward scattering direction).  If we consider a limit dominated by point particle $t$-channel exchange, the factor $\Gamma(-1-\alpha't)$ reduces to a particle pole $1/t$.  If we assumed that this corresponds to $A_{\text{point}}$ above, we could attempt to Fourier transform the remainder of the amplitude to obtain the charge density $\rho(\vec x_\perp)$.  However, this procedure yields a complex function (taking a single transverse direction for simplicity)
\be\label{transFT}
\int dq_\perp\, s^{-\alpha' q_\perp^2} e^{i\pi \alpha'q_\perp^2} e^{i q_\perp b}=\frac{\exp\left(-\frac{b^2}{4 \alpha' (\log s-i \pi )}\right)}{\sqrt{\frac{\log s}{\pi }-i}},
\ee
rather than a real charge distribution of the form $|\Psi|^2$.  As such, it has no immediate interpretation in terms of the transverse distribution of the source string density.  Note that this complex quantity is also not directly the incoming wavefunction $\Psi$ of either string; in light cone gauge this is simply given in (4.20) of \cite{bpst}.       

One could conjecture (as may have been done in \cite{Giddings})\footnote{Although we stress that we do not presume to express the views of others.} that the complex quantity (\ref{transFT}) is this wavefunction (as a function of the spreading $\Delta y$ of the endpoint of the string instead of $b$), but with an appropriate cutoff on the source string modes obtained by measuring it with the other string.  By construction, the Fourier transform of the quantity (\ref{transFT}) with respect to $b$ is equal to the scattering amplitude, so identifying (\ref{transFT}) with the wavefunction of one of the incoming states potentially enables a transverse interpretation of the four-point amplitude discussed in Section \ref{longvstrans}.    

We will contrast this with the wavefunction discussed in the main text, and below, which results from the free string wavefunction simply cut off at a mode number $n_{\text{max}}$.  
The calculation in \cite{lennyspreading}\ shows that the light cone gauge wave function, cut off at a mode number $n_{\text{max}}$ is a Gaussian function of the transverse spreading $\Delta X^+$, with a width $\sim\sqrt{\alpha'\log n_{\text{max}}}$.  In the physical picture of \cite{lennyspreading}, the value of $n_{\text{max}}$ is determined by the light cone time resolution, giving (\ref{nmax}) as reviewed in detail in \cite{usBH}.  
The conjecture that (\ref{transFT}) is the correct cut-off wavefunction instead would formally correspond to the cutoff $n_{\text{max}}$ in this Gaussian wavefunction being replaced by $e^{-i\pi}n_{\text{max}}$.  But the real positive cutoff $n_{\text{max}}$ (\ref{nmax}) appears explicitly in the light cone computation, in the sum over mode number $n$ in (4.20) of \cite{bpst}.   This positive cutoff on  mode number $n$, combined with the Gaussian statistics of the free transverse worldsheet fields leads to a wavefunction like (\ref{transFT}) but without the $-i\pi$ offset. 

Let us elaborate on this to be very clear.  We can contrast the wavefunction
\be\label{Psizero}
\Psi_0(\Delta y) = N_0 \exp\left(-\frac{(\Delta y)^2}{\alpha' \log(n_{\text{max}}/c)}\right)
\ee
(for real positive $c$)   from the above Fourier transform
 \be\label{Psipi}
 \Psi_\pi(\Delta y) = N_\pi \exp\left(-\frac{(\Delta y)^2}{\alpha'(\log n_{\text{max}}-i\pi)}\right)
 \ee
 Both of these wavefunctions produce an expectation value $\langle( \Delta y)^2\rangle \sim \log n_{\text{max}} $ at leading order at large $\log n_{\text{max}}$.  The first wavefunction (\ref{Psizero}) has no corrections to this aside from the $n_{\text{max}}$-independent shift $-\log c$, whereas the second wavefunction leads to a distinctive correction term
 \be\label{Psipiexp}
 \langle \Psi_\pi|( \Delta y)^2 |\Psi_\pi \rangle \sim \log n_{\text{max}}-\frac{\pi^2}{\log n_{\text{max}} }.
 \ee  
 This second term does not arise in the mode sum 
 \be\label{transversesum}
 \langle(\Delta X_\perp)^2\rangle =\sum_{n=1}^{n_{\text{max}}}\frac{1}{n} 
 \ee
appearing in \cite{lennyspreading}\ with a simple cutoff $n_{\text{max}}$,  even including subleading effects at large $n_{\text{max}}$.  To see this, apply the Euler-MacLaurin formula
 \be\label{EMac}
\sum_{n=a}^b f(n)=\int_a^b f(n)dn+\frac{1}{2}(f(b)+f(a))+\sum_{j=1}^\infty (-1)^j \frac{B_j}{(2j)!}(f^{(2j-1)}(a) -f^{(2j-1)}(b)),
 \ee
 which gives for our sum (\ref{transversesum})
 \be\label{sumapprox}
 \int_1^{n_{\text{max}}}\frac{dn}{n}+\frac{1}{2}+\sum_{j=1}^\infty\frac{ (-1)^j}{2j}B_j+O(1/n_{\text{max}}).
 \ee
 The first term gives precisely the leading $\log n_{\text{max}} $ term, and the remainder contains no contribution depending on $n_{\text{max}}$ like the offset $\pi^2/\log n_{\text{max}}$ term in (\ref{Psipiexp}).  (The constant remainder here can easily be absorbed via the constant $c$ in the  wavefunction $\Psi_0$.)
 Altogether, the wavefunction $\Psi_0$ is consistent with \cite{lennyspreading}, with the simple cutoff $n_{\text{max}}$ in (\ref{transversesum}).  The alternative (\ref{Psipi}) is not consistent with this.   
 
 That said, it is a logical possibility that a more complicated measurement process could introduce this imaginary offset into the effective wavefunction.  We are aware of no derivation of this alternative hypothesis;  instead the appearance of the predicted $n_{\text{max}}$ in \cite{bpst}\ provides concrete evidence in favor of the simple picture \cite{lennyspreading}\ giving (\ref{transversesum}).   

Another alternative one might entertain is to keep $n_{\text{max}}$ positive in the Gaussian wavefunction \cite{lennyspreading}, but change its value.  In particular, in order to capture the peak impact parameter derived in the main text,  it would need to be the case that the measurement depends crucially on the scattering angle $\theta$, via $t\sim -E^2\theta^2$.  Again, the cutoff on mode number manifest in (4.20) of \cite{bpst}\ does not exhibit such dependence.   

Rather than make either alternative conjecture in the previous two paragraphs, we find it physically much more plausible to include in our analysis the assumption that the transverse string spreading is given by \cite{lennyspreading}, for the reasons just reviewed.  Given that, it is possible to determine whether an amplitude is proceeding via a coupling at the tail of this transverse distribution by checking for the corresponding multiplicative suppression factor as discussed and applied in the main text.

\begingroup\raggedright\begin{thebibliography}{10}

\baselineskip=14.5pt
\bibitem{lennyspreading}
L. Susskind,
``Strings, black holes and Lorentz contraction,"
Phys. Rev. D {\bf 49}, 6606-6611 (1994). \\ 
M. Karliner, I. R. Klebanov, and L. Susskind,
``Size and Shape of Strings," 
Int. J. Mod. Phys. {\bf A3} 1981 (1988). 
[hep-th/9308139].
\bibitem{bpst}
R. C. Brower, J. Polchinski, M. J. Strassler, C. Tan,
``The Pomeron and gauge/string duality,"
JHEP {\bf{0712}} 005 (2007) [hep-th/0603115].
\bibitem{usBH}
M. Dodelson and E. Silverstein, ``String-theoretic breakdown of effective field theory near black hole horizons," to appear.  
\bibitem{locality}
D. Lowe, J. Polchinski, L. Susskind, L. Thorlacius, J. Uglum,
``Black hole complementarity versus locality,"
Phys. Rev. D {\bf{52}} 6997-7010 (1995) [hep-th/9506138].
\bibitem{JoeBHcomp} 
  J.~Polchinski,
  ``String theory and black hole complementarity,''
  In *Los Angeles 1995, Future perspectives in string theory* 417-426
  [hep-th/9507094].
\bibitem{ggm}
S. B. Giddings, D. J. Gross, and A. Maharana,
``Gravitational effects in ultrahigh-energy string scattering,"
Phys. Rev. D {\bf 77} 046001 (2008) [hep-th/0705.1816].

\bibitem{firewalls}
A. Almheiri, D. Marolf, J. Polchinski, and J. Sully,
``Black Holes: Complementarity or Firewalls?"
JHEP {\bf 062} 1302 (2013) [hep-th/1207.3123].
\bibitem{backdraft}
E. Silverstein,
``Backdraft: String Creation in an Old Schwarzschild Black Hole," (2014) [hep-th/1402.1486].

\bibitem{Veneziano}
G.~Veneziano,
  ``Construction of a crossing - symmetric, Regge behaved amplitude for linearly rising trajectories,''
  Nuovo Cim.\ A {\bf 57}, 190 (1968).

\bibitem{yoyo1}
 W. A. Bardeen, I. Bars, A. J. Hanson, and R. D. Peccei, ``A Study of the Longitudinal
Kink Modes of the String," Phys. Rev. D13 (1976) 2364-2382.\\ 
\bibitem{yoyo2}
A. Ficnar and S. S. Gubser, ``Finite momentum at string endpoints," Phys. Rev. D89 (2014) [hep-th/1306.6648].
\bibitem{yoyo3}
X. Artru, ``Classical String Phenomenology. 1. How Strings Work," Phys. Rept. 97 (1983) 147.
\bibitem{juancausality}
 X.~O.~Camanho, J.~D.~Edelstein, J.~Maldacena and A.~Zhiboedov,
  ``Causality Constraints on Corrections to the Graviton Three-Point Coupling,''
  arXiv:1407.5597 [hep-th].

\bibitem{GSW}
M.~B.~Green, J.~H.~Schwarz and E.~Witten,
  ``Superstring Theory. Vol. 1: Introduction,''
   Cambridge Monogr.Math.Phys..
 
   M.~B.~Green, J.~H.~Schwarz and E.~Witten,
  ``Superstring Theory. Vol. 2: Loop Amplitudes, Anomalies And Phenomenology,''
  Cambridge, Uk: Univ. Pr. ( 1987) 596 P. ( Cambridge Monographs On Mathematical Physics)
\bibitem{sst}
N. Seiberg, L. Susskind, and N. Toumbas,
``Space/Time Non-Commutativity and Causality,"
JHEP {\bf{0006}} 044 (2000)  [hep-th/0005015].
\bibitem{grossmende}
 D.~J.~Gross and P.~F.~Mende,
  ``String Theory Beyond the Planck Scale,''
  Nucl.\ Phys.\ B {\bf 303}, 407 (1988).
\bibitem{BR}
K.~Bardakci and H.~Ruegg,
  ``Meson resonance couplings in a five-point veneziano model,''
  Phys.\ Lett.\ B {\bf 28}, 671 (1969).

\bibitem{BP}
A. Bialas and S. Pokorski, ``High Energy Behavior of the Bardakci-Ruegg Amplitude"
\bibitem{JoeBook}

J.~Polchinski,
  ``String theory. Vol. 1: An introduction to the bosonic string,''
  Cambridge, UK: Univ. Pr. (1998) 402 p

\bibitem{zak}
W. J. Zakrzewski, ``Signature in the Bardakci and Ruegg model," Nucl. Phys. B {\bf 14}, 458 (1969).

\bibitem{strassler5pt}
C. P. Herzog, S. Paik, M. J. Strassler, E. G. Thompson, 
``Holographic Double Diffractive Scattering," JHEP 0808 (2008) 010 [hep-th/08060181].

\bibitem{formfactors}
http://www.nist.gov/data/PDFfiles/jpcrd67.pdf;
http://www.cbooth.staff.shef.ac.uk/phy304/scattering.html
\bibitem{Giddings}

S. Giddings, private discussions


\endgroup
\end{document}